\documentclass{aa}  
\usepackage{graphicx}
\usepackage{txfonts}
\usepackage{mathabx}
\begin{document} 

   \title{Enrichment of the HR~8799 planets by minor bodies and dust}

   \author{K.~Frantseva\inst{1,2},
           M.~Mueller\inst{3,4,2},
           P. Pokorn\'{y}\inst{5,6},
           F.~F.S.~van~der~Tak\inst{2,1},
           I.~L.~ten~Kate\inst{7} }

\institute{Kapteyn Astronomical Institute, University of Groningen, Landleven 12, 9747 AD Groningen, The Netherlands \\ \email{frantseva@astro.rug.nl} 
\and SRON Netherlands Institute for Space Research, Landleven 12, 9747 AD Groningen, The Netherlands
\and NOVA Netherlands Research School for Astronomy, The Netherlands
\and Leiden Observatory, Leiden University, Leiden, The Netherlands
\and Department of Physics, The Catholic University of America, Washington, DC, 20064, USA,
\and Astrophysics Science Division, NASA Goddard Space Flight Center, Greenbelt, MD, 20771, USA
\and Department of Earth Sciences, Utrecht University, Budapestlaan 4, 3584 CD Utrecht, The Netherlands}

   \date{Received 25 September 2019; accepted 13 May 2020}

\titlerunning{Enrichment of the HR~8799 planets by asteroids and comets}
\authorrunning{K.~Frantseva et al.}

  \abstract
   {In the Solar System, minor bodies and dust deliver various materials to planetary surfaces. Several exoplanetary systems are known to host inner and outer belts, analogues of the main asteroid belt and the Kuiper belt, respectively.}
   {We study the possibility that exominor bodies and exodust deliver volatiles and refractories to the exoplanets in the well-characterised system HR~8799.}
   {We performed N-body simulations to study the impact rates of minor bodies in the system HR~8799. The model consists of the host star, four giant planets (HR~8799~e, d, c, and b), 650,000 test particles representing the inner belt, and 1,450,000 test particles representing the outer belt. Moreover we modelled dust populations that originate from both belts.}
   {Within a million years, the two belts evolve towards the expected dynamical structure (also derived in other works), where mean-motion resonances with the planets carve the analogues of Kirkwood gaps. We find that, after this point, the planets suffer impacts by objects from the inner and outer belt at rates that are essentially constant with time, while dust populations do not contribute significantly to the delivery process. We convert the impact rates to volatile and refractory delivery rates using our best estimates of the total mass contained in the belts and their volatile and refractory content. Over their lifetime, the four giant planets receive between $10^{-4}$ and $10^{-3}\,M_\Earth$ of material from both belts.}
   {The total amount of delivered volatiles and refractories, ${5\times10^{-3}\,\textrm{M}_\Earth}$, is small compared to the total mass of the planets, $11\times10^{3}\,\textrm{M}_\Earth$. However, if the planets were formed to be volatile-rich, their exogenous enrichment in refractory material may well be significant and observable, for example with JWST-MIRI. If terrestrial planets exist within the snow line of the system, volatile delivery would be an important astrobiological mechanism and may be observable as atmospheric trace gases.}
   

   \maketitle

%-------------------------------------------------------------------

\section{Introduction}\label{sec5.1:Introduction}

The first exoplanet discovery \citep{Mayor1995} around a main-sequence star started a new chapter in the field of planetary sciences. More than 4,000 confirmed exoplanets have been detected as of May 2020 and the number doubles every $\sim$27 months\footnote{\url{http://exoplanet.eu}}. The variety of the discovered exoplanets shows that planetary systems are very common (planets are probably as common as stars) and diverse \citep{Perryman2018}. There are exoplanets with masses ranging from the Moon's mass to hundreds of Jupiter masses, with orbital periods ranging from hours to thousands of years and with a great variety in mass densities from that of  H$_2$ gas to that of solid iron \citep{Winn2015}. A large fraction of the discovered exoplanets do not have analogues in the Solar System; this is the case for super-Earths, with masses between about 1 and 10 Earth masses, and hot Jupiters, giant planets orbiting very close to their host stars \citep{Raymond2018}, for example. Importantly, though, there appears to be a population of terrestrial planets at distances to the star that may allow liquid water to exist on the surface, possibly allowing life as we know it to form.

\begin{figure*}
\centering
{\includegraphics[width=.49\linewidth]{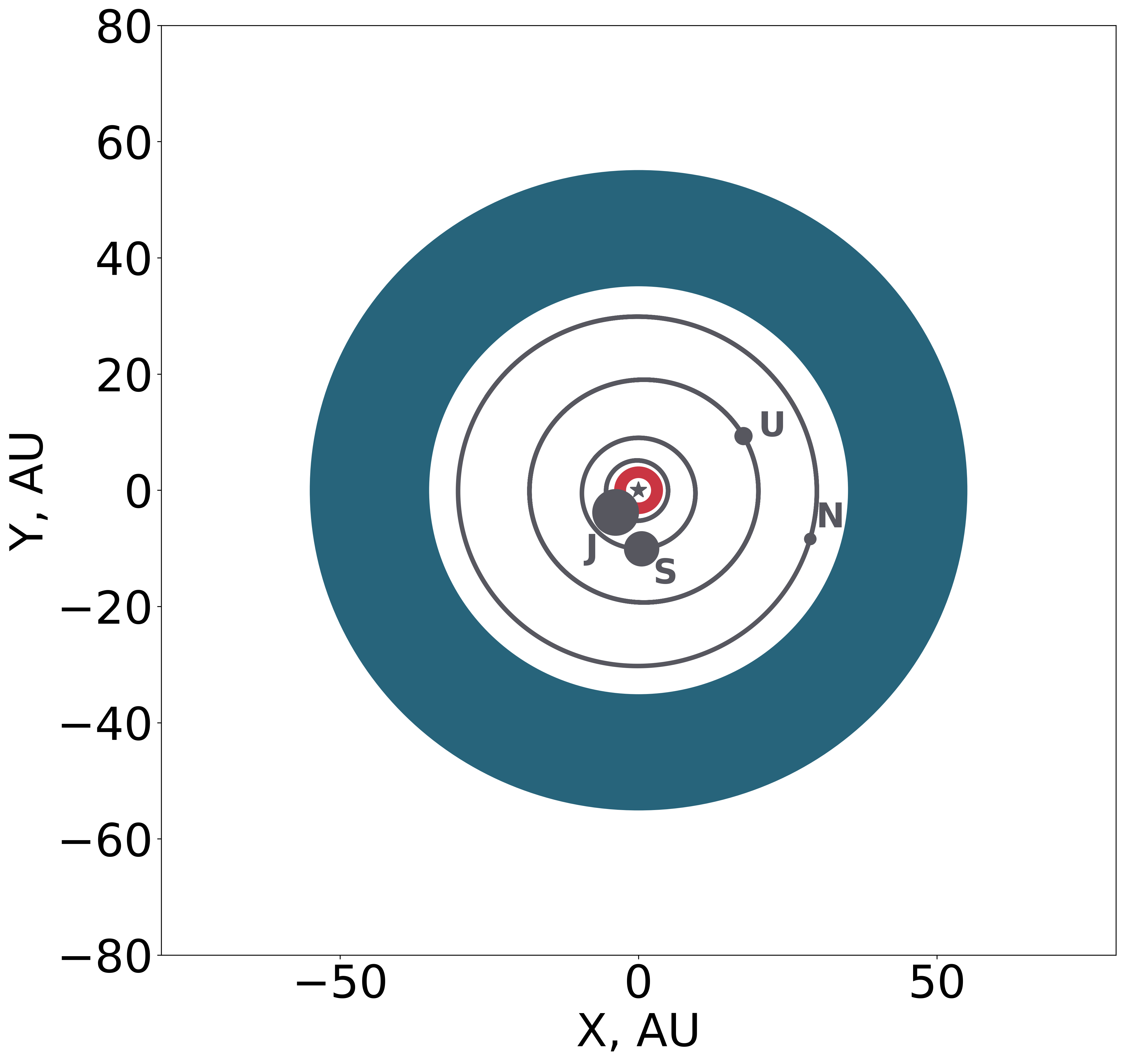}}
\hspace{\fill}
{\includegraphics[width=.49\linewidth]{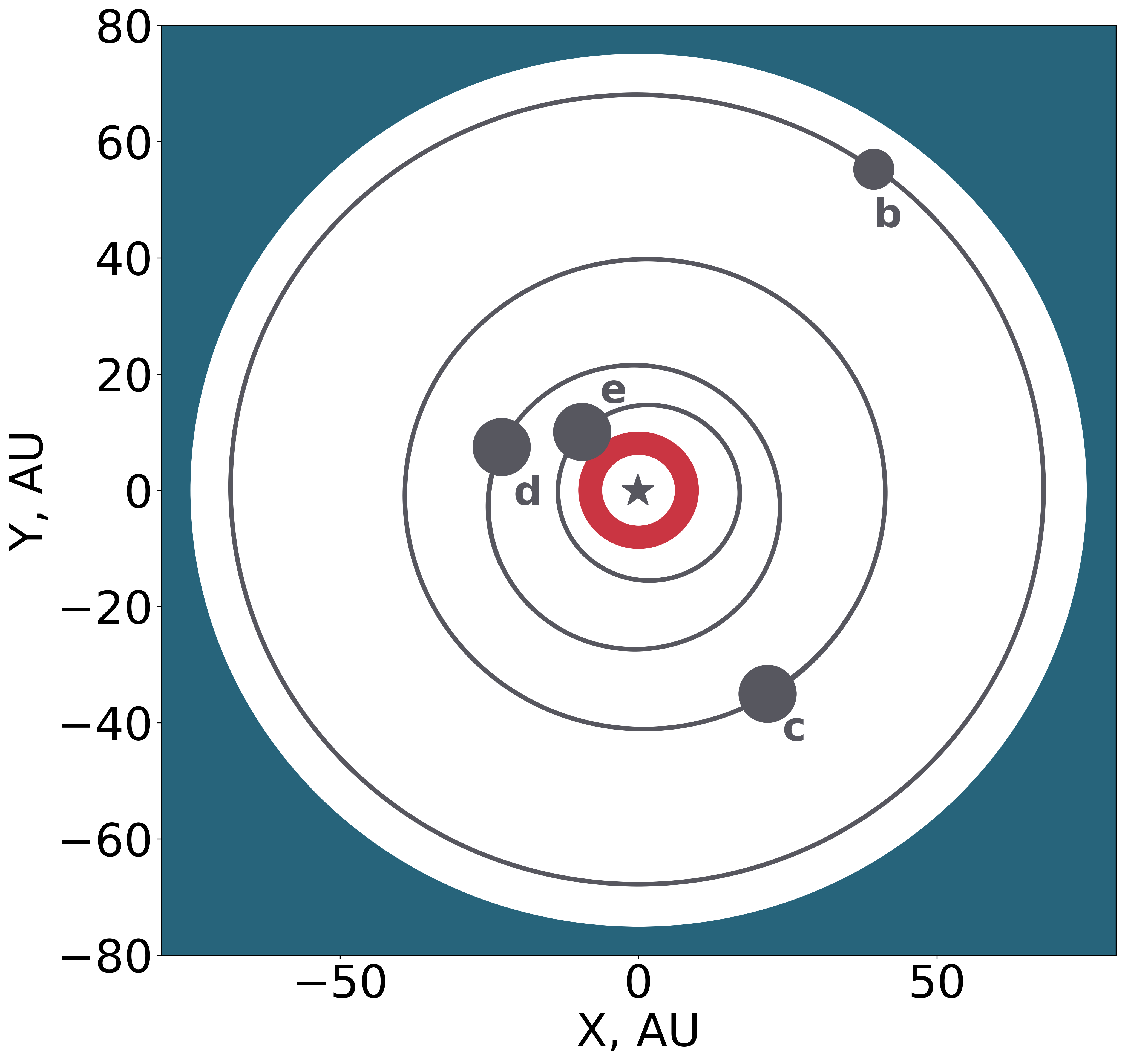}}
\caption{Orbital structure of the Solar System, left, and the HR~8799 system, right. The inner belts are represented in red and the outer belts in blue. Planet masses are indicated by the size of the dots; each plot has a separate log scale. We note that the orbital eccentricities of the giant planets in the HR~8799 system are several times larger than the eccentricities of the Solar System giants, which lead to more impacts between planets and small bodies.}
\label{fig5.1:1}
\end{figure*}

Apart from looking for analogues of the Solar System planets, a search for analogues of small body populations is ongoing. Recent observations \citep{Acke2012,Booth2016,Close2010,Lagrange2010,Matthews2014,Moor2013,Su2013,Welsh2013} have found analogues of the main asteroid belt (MAB) and the Kuiper belt (KB) around $\sim$20\% of the nearest stars \citep{Ren2019}. Within our own Solar System, the small body belts (MAB and KB) play an important role in delivering volatiles to planets \citep[e.g.][]{Frantseva2018,Frantseva2020}. 

Delivery is one of the two major processes that has shaped the atmospheres of the terrestrial planets in our Solar System, with outgassing being the other \citep{DePater2015, Smallwood2018, Chen2019}. In the MAB, certain types of asteroids, C-types, together with comets and small bodies from the KB are known to contain significant amounts of water and organic compounds. The migration of these small bodies occasionally results in impacts with planets, which in turn leads to volatile enrichment of the planets. The existence of asteroid belts in other planetary systems implies that the same volatile delivery mechanisms might be at play around other stars. This way, volatile material, in particular water and organic compounds, can be delivered to exoplanets. Studying these delivery mechanisms is of astrobiological interest and will lead to better understanding of planetary formation, evolution, and habitability \citep{Raymond2018}. Enrichment of exoplanet atmospheres in refractory materials is of less astrobiological relevance, but may become observable using the Mid-InfraRed Instrument aboard the James Webb Space Telescope (JWST-MIRI) due to be launched in 2021. Silicate features in particular feature prominently in MIRI's wavelength range \citep[e.g.][]{Rieke2015}. The atmospheres of exoplanets beyond the snow line would not be expected to contain significant amounts of refractory materials, so any detection thereof would be diagnostic of impact-driven enrichment.

In this paper we focus on the exoplanetary system HR~8799, which is known to host both a warm and a cold debris belt, respective analogues of the Main Asteroid Belt and the Kuiper Belt. Modelling such a planetary system helps in understanding the interaction between planets and planetary debris. 

HR~8799 is a nearby young A5V star with mass ${\approx1.56\,\textrm{M}_\odot}$ \citep{Gray1999,Gray2003,Baines2012,Gozdziewski2014}. The exact age of the star is somewhat uncertain between 30 and 60~Myr \citep{Marois2010,Zuckerman2011}. HR~8799 is the host star of four giant planets HR~8799~e, HR~8799~d, HR~8799~c, and HR~8799~b detected via direct imaging \citep{Marois2008,Marois2010}. HR~8799~e, HR~8799~d and HR~8799~c have masses of approximately 9M$_\textrm{jup}$, and HR~8799~b of 7M$_\textrm{jup}$. For comparison, the giant planets in the Solar System have masses of 1M$_\textrm{jup}$, 0.3M$_\textrm{jup}$, 0.05M$_\textrm{jup}$ and 0.04M$_\textrm{jup}$, see Figure~\ref{fig5.1:1}. HR~8799~e, HR~8799~d, HR~8799~c and HR~8799~b orbit their host star at distances of 15 - 69~AU, see Table~\ref{tbl5.2:1} for more details. The Solar System giant planets orbit closer to the host star at distances of 5 - 30~AU.

The orbits of the HR~8799 planets are not well characterised observationally, yet. The available astrometry spans only $\simeq$ 12\% of the innermost planet's orbit and $\simeq$ 3\% of the outermost planet's. A variety of orbital configurations is consistent with the data, most of which are dynamically unstable; however, a few stable configurations were found \citep{Gozdziewski2014,Gotberg2016,Gozdziewski2018,Wang2018}. The best long-term stable solution assumes that the four planets are locked in a protective mean motion resonance 1:2:4:8 \citep{Gozdziewski2014,Gozdziewski2018}. The higher planetary masses relative to the star as well as the larger eccentricities make planet-planet interactions more important in HR~8799 relative to the Solar System. This leads to many unstable orbital configurations consistent with the data (see above), and also requires extra care in numerical modelling. Detection of double Laplace resonances or higher-order resonant chains, such as proposed for the HR~8799 planets or is known for the Galilean satellites \citep{Sinclair1975,Peale2002}, is demanding \citep{Michtchenko2001,Gallardo2016}. A good understanding of the orbital architecture of the system is essential for studying the system's debris belts that have been observed with the Herschel Space Observatory, the Atacama Large Millimetre/submillimetre Array (ALMA) and the Spitzer Space Telescope \citep{Su2009,Su2014,Matthews2014,Booth2016}. 

The system contains at least two distinct debris belts: an inner belt and an outer belt. The outer belt was discovered in imaging with the Multi-Band Imaging Photometer for Spitzer (Spitzer-MIPS) and is spatially resolved, while the inner belt is too small to be spatially resolved, but was inferred from the observed infrared (IR) excess in the spectral energy distribution (SED) using the Spitzer Infrared Spectrograph (Sptizer-IRS) data \citep{Su2009}. The inner belt lies between $\sim$6~AU and $\sim$15~AU, and the outer belt extends from $\sim$90~AU to $\sim$310~AU. The inner belt lower dust mass limit was estimated to be $1.1 \times 10^{-6} \textrm{M}_\Earth$ and the outer belt's is $0.12 \textrm{M}_\Earth$ \citep{Su2009}. These estimated belt masses are dust-only masses, while large objects (if they exist) would by far dominate the mass budget. The structure of HR~8799 resembles the structure of our own Solar System but on a larger scale as shown in Fig.~\ref{fig5.1:1}. The Solar System up to the Kuiper belt can be fitted inside the orbit of the outermost planet of the HR~8799 system. Both systems contain an inner warm belt close to the host star, within $\sim$6~AU and $\sim$15~AU for  HR~8799 and within $\sim$2~AU and $\sim$4~AU for the Solar System. Furthermore, both systems contain an outer cold belt, within $\sim$90~AU to $\sim$310~AU for  HR~8799 and $\sim$35~AU to $\sim$100~AU for the Solar System. Between the two belts each of the systems have four giant planets. Remarkably, HR~8799 and the Solar System show the same overall structure although the HR~8799 system is much larger than the Solar System and its planets are much more massive (see Fig.~\ref{fig5.1:1}). We note that terrestrial planets may exist inside the warm belt of HR~8799; those would not be detectable using current telescopes. 

\begin{table*}
\caption{Orbital elements of 4 known planets HR~8799 e,d,c and b as in \citet[Table 1]{Gozdziewski2014}. In our simulations we adopted the nominal values from the table.  $M_\textrm{p}$, $a$, $e$, $i$, $\Omega$, $\omega$ and $M$ correspond to the mass of the planet in Jupiter masses, the semi-major axis of the planet in AU, the eccentricity, the inclination in degrees, the longitude of the ascending node, the argument of pericentre and the mean anomaly at the epoch 1998.83 respectively. $i$ is the inclination of coplanar orbits to the sky plane. $i$ and $\Omega$ are assumed to be identical for all four planets; $i$ is the best-fitting inclination of coplanar orbits to the sky plane. The stellar mass $M_*$ is 1.56M$_{\odot}$.}
\label{tbl5.2:1}
\centering
\begin{tabular}{crrrrrrr}
\hline
planet & $M_\textrm{p}$ (M$_\textrm{jup}$) & $a$ (AU) &  $e$ &  $i$ (deg) &  $\Omega$ (deg) &  $\omega$ (deg) &  $M$ (deg) \\
\hline
e & 9 $\pm$ 2 & 15.4 $\pm$ 0.2 & 0.13 $\pm$ 0.03 & 25 $\pm$ 3 & 64 $\pm$ 3 & 176 $\pm$ 3 & 326 $\pm$ 5  \\
d & 9 $\pm$ 3 & 25.4 $\pm$ 0.3 & 0.12 $\pm$ 0.02 & 25 $\pm$ 3 & 64 $\pm$ 3 & 91 $\pm$ 3 & 58 $\pm$ 3 \\
c & 9 $\pm$ 3 & 39.4 $\pm$ 0.3 & 0.05 $\pm$ 0.02 & 25 $\pm$ 3 & 64 $\pm$ 3 & 151 $\pm$ 6 & 148 $\pm$ 6 \\
b & 7 $\pm$ 2 & 69.1 $\pm$ 0.2 & 0.020 $\pm$ 0.003 & 25 $\pm$ 3 & 64 $\pm$ 3 & 95 $\pm$ 10 & 321 $\pm$ 10 \\
& & & & & & & \\
\hline
\end{tabular}
\end{table*}

The interaction between the giant planets of HR~8799 and the debris belts have been studied by \citet{Contro2015,Contro2016,Read2018,Gozdziewski2018}. Numerical simulations by \citet{Contro2016} indicate that the inner belt is structured, with gaps at the locations of mean motion resonances with the innermost planet similar to the Kirkwood gaps in the MAB. Moreover, the simulations suggest that the belt is located between 6~AU and 8~AU and collisions within the belt occur at velocities of the order of 1.2~km/s or less, lower than the collisional velocities in the MAB, which are of order 5~km/s. We note that \citeauthor{Contro2016} suggest that the outer edge of the inner belt has to be located at 8~AU, and not 15~AU as suggested by past observations by Spitzer \citep{Su2009}. The simulations demonstrate that due to the influence of the innermost planet it is not possible to keep a population of small bodies between 8 and 15~AU. Similarly, \citet{Read2018} performed N-body simulations to study the effect of the giant planets on the outer belt and demonstrated that the belt has a similar structure as the inner belt, with gaps at the locations of the mean motion resonances with the outermost planet. \citet{Read2018} show that adding a hypothetical planet with mass $0.1\,-\,1$ M$_\textrm{jup}$ outside HR~8799~b, the outermost confirmed planet, pushes the belt  outwards, providing a better fit to ALMA observations \citep{Booth2016,Read2018,Gozdziewski2018}. Such a planet cannot be detected with current telescopes.

In our own Solar System the interaction between the belts and the planets leads to scattering of small bodies within the system and beyond. Sometimes this scattering leads to impacts with the planets. For the terrestrial planets impacts can appreciably enrich the surface in various materials including volatiles. For the giant planets refractory enrichment is even observable, as in the case of the Shoemaker-Levy~9 impacts in 1994 \citep{ATREYA1999,Harrington2004,Fletcher2010}. Here, we test for the first time how strong this effect is in the HR~8799 system. We perform several sets of dynamical N-body simulations with 1,600,000 test particles representing exoasteroids and exocomets. These simulations result in minor body impact rates with the giant planets that are then converted to volatile delivery rates. We also perform a set of N-body simulations of dust to analyse a potential mass transport from the inner and outer dust belts onto the HR~8799 planets.

In Section~\ref{sec5.2:NumericalSimulations} we describe our N-body simulations. Section~\ref{sec5.3:SimulationResults} presents the results of the numerical simulations, checks for consistency with previous results, and determines impact rates.  In Section~\ref{sec5.4:VolatileDeliveryRates} we convert the latter to volatile delivery rates to the planets. Discussion and conclusions are presented in Sections~\ref{sec5.5:Discussions} and \ref{sec5.6:Conclusions}.

%-------------------------------------------------------------------

\section{Numerical simulations}\label{sec5.2:NumericalSimulations}

We have performed numerical simulations of the dynamical evolution of dust and minor bodies in the inner and outer belts in the exoplanetary system HR~8799 to study the volatile and refractory influx onto the four known giant exoplanets.

%-------------------------------------------------------------------

\subsection{Minor bodies}

To model the minor body populations of the inner and outer belts, we used the $N$-body integrator MERCURIUS \citep{Rein2019} from the software package REBOUND \citep{Rein2012}. We set up a model of the motion of one gravitationally dominant object (HR~8799; ${M_\star=1.56\,\textrm{M}_\odot}$) and $N$ massive objects (in our case: $N=4$ representing the four planets in the system) under the influence of their mutual gravity over Myr timescales forward in time. Simulations focusing on the inner and outer belts were done separately, as described in \S\ref{sec5.2.1:NumericalSimulationsInnerBelt} and \S\ref{sec5.2.2:NumericalSimulationsOuterBelt}. The inner belt is represented by 650,000 test particles and the outer belt by 1,450,000. Our first run of the outer belt simulation contained 650,000 test particles and one of the planets received 0 impacts. To overcome small number statistics we added more test particles to the outer belt simulation. To achieve a $3\sigma$ result while adopting $\sqrt{N}$, where $N$ is the number of impacts, for the noise we need $N=10$ as a minimum. Therefore we require the total number of impactors on each planet to be always well above 10. Increasing the number of test particles resulted in total impacts on each planet to be more than 10 and helped us to achieve a $3\sigma$ result.

In our simulation, the test particles move passively under the influence of the combined gravitational potential of the star and the planets. The gravitational effect of the test particles on one another and on the massive objects is neglected. Test particles are removed from the simulation once they collide with a planet or with the star. Test particles are considered ejected from the planetary system and discarded when they exceed a belt-dependent heliocentric distance: 1,000~AU for the inner-belt simulations and 10,000~AU for the outer-belt simulations following \citet{Contro2016} and \citet{Read2018}, respectively. Timestep values were set separately for the inner and outer belts, see \S\ref{sec5.2.1:NumericalSimulationsInnerBelt} and \S\ref{sec5.2.2:NumericalSimulationsOuterBelt} for a detailed description. The simulation results are recorded by taking snapshots of the simulations at predefined times (6~Myr, 30~Myr, 60~Myr and 70~Myr).

Table~\ref{tbl5.2:1} presents the planetary initial conditions that have been adopted for our simulations. The masses and orbital parameters were taken from \citet[their Table 1]{Gozdziewski2014}. The inclination, i, of the four planets is the best-fitting inclination of coplanar orbits to the sky plane, while the inclination of the HR~8799 equator to the sky plane is $\sim23$ degrees as follows from the statistical analysis of the rotational speed of A5 stars. For the longitude of the ascending node, $\Omega$, \citet{Gozdziewski2014} assumed the same value for all four planets. The planetary radii (0.0005327~AU, 0.0005327~AU, 0.0005327~AU and 0.000538629~AU) were calculated using the "Jovian Worlds" mass-to-radius relation from \citet{Chen2017}:

\begin{equation}
\frac{R}{\textrm{R}_\Earth} = 17.74 \, \left(\frac{M}{\textrm{M}_\Earth}\right)^{-0.044},
\end{equation}

which is valid for $0.414 \, \textrm{M}_\textrm{jup}
\, < \, M \, < \, 83.79 \, \textrm{M}_\textrm{jup}$. The relation represents a population of planets with masses larger than a sufficient mass for gravitational self-compression that starts reversing the growth of the planet (therefore the exponent is negative). Radii are needed to model impacts: a test particle is considered to have impacted the star or a planet once it ventures within its radius. For the radius of the star, we adopt a value of $1.440\,\pm\,0.006\,\textrm R_{\odot}$ \citep{Baines2012} and 60~Myr for its age \citep{Marois2010}. 

%__________________________________________________________________

\subsubsection{Inner belt}\label{sec5.2.1:NumericalSimulationsInnerBelt}

To model the inner belt of the HR~8799 we followed the initial conditions described in \citet{Contro2016}. We used 650,000 test particles \citep[500,000 in][]{Contro2016} with semi-major axes between 1 and 10~AU. 

We adopt 1~AU as the inner edge value of the simulations following \citet{Contro2016}. The outer boundary, 10~AU, follows the estimated observational value as in \citet{Marois2010}. The eccentricities of the particles are distributed uniformly between 0 and 0.1 and the inclinations are set between 0$^\circ$ and 5$^\circ$. The remaining, angular, orbital elements are set to random numbers between 0$^\circ$ and 360$^\circ$. The simulations were performed for 70~Myr forward in time. The timestep of the simulations is set to 7~days in order to resolve the orbits of the innermost test particles orbiting the star at 1~AU with an orbital period of 292~days.

%__________________________________________________________________

\subsubsection{Outer belt}\label{sec5.2.2:NumericalSimulationsOuterBelt}

The outer belt simulations were based on the initial conditions from \citet{Read2018}. We created 1,450,000 test particles \citep[50,000 in ][]{Read2018} with semi-major axes between 69.1 and 429~AU. The inner boundary of the outer belt is set to the semi-major axis of the outermost planet HR~8799~b. We note that ALMA observations \citep{Booth2016} suggest that the inner edge of the disc has to be further out at 145~AU, which might be explained by an additional fifth planet. However, we stick to the initial conditions described in \citet{Read2018} for consistency. The outer edge of the disc follows from the ALMA observations by \citet{Booth2016}. Eccentricities were set between 0 and 0.05 and inclinations between 0$^\degree$ and 2.86$^\degree$ (corresponding to 0.05~rad). The remaining, angular, orbital elements are randomly distributed between 0$^\degree$ and 360$^\degree$ as for the inner belt. The simulations were performed forward in time for 70~Myr. The simulation's timestep was set to 0.48~yr in order to resolve the orbit of the innermost planet HR~8799~e, which has an orbital period of 48~years, and any close encounters that would occur between the planet and test particles.

%-------------------------------------------------------------------

\subsection{Dust}

We model dust originating from the inner and outer belts. In this manuscript, we call all particles smaller than minor bodies collectively dust, even though, by definition, dust larger than 30 micrometres in diameter should be called meteoroids \footnote{\url{https://www.iau.org/static/science/scientific_bodies/commissions/f1/meteordefinitions_approved.pdf}}. 

For the source body populations we use the orbital element distributions of the small bodies in the inner and outer belts at time 70~Myr from our N-body simulation, see \S\ref{sec5.2.1:NumericalSimulationsInnerBelt} and \S\ref{sec5.2.2:NumericalSimulationsOuterBelt}. This is because the dust populations in HR~8799 based on our simulations are not dynamically stable for more than 40~Myr and the parent body configuration is quite similar between 30~Myr and 70~Myr. For the inner belt we released 875,000 particles with diameters between 1.5 and 4.5 $\mu$m, uniformly distributed over 7 bins. For the outer belt we released 17,500 particles with diameters between 10 and 1000 $\mu$m, also uniformly distributed over 7 bins. These size distributions are chosen based on the Spitzer, Herschel and ALMA observations of dust grains in the inner and outer belts \citep{Su2009,Matthews2014,Booth2016}. The difference between the number of particles in both belts is caused by the difference in the computational power needed to complete the dynamical evolution of the dust particles: 50~Myr for the outer belt versus several Myr for the inner belt. We assume a dust bulk density of $\rho=2$~g~cm$^{-3}$. 

All particles were numerically integrated under the gravitational influence of four planets, the host star, and the effects of radiation pressure and Poynting-Robertson (PR) drag assuming a stellar luminosity of $L_\star=4.92\,\textrm{L}_\odot$ \citep{Gray_Kaye_1999} using the \texttt{SWIFT\_RMVS\_3} numerical integrator \citep{Levison1994}. Particles were tracked until they vanished in one of the sinks: too close to the host star ($R<0.05$~AU), too far from the host star ($R>1,000$~AU), or impact on one of the four planets. No particle in our simulation was stable for more than 70~Myr, that is, the age of the system. Particle orbital elements were recorded every 1,000 years, where each recorded set of orbital elements is assumed to be a unique representation of meteoroids. This in other words means that we assume that the dust population is in a steady-state and dust is generated every 1,000 years to feed the dynamically evolving dust cloud. This results in a simulation containing 875,000 and 15,000 dynamical pathways for the inner and outer belt dust populations respectively, that is sampled by millions of unique dust orbital configurations.

%__________________________________________________________________

\section{Simulation results: Orbital structure, impact rates}\label{sec5.3:SimulationResults}

%__________________________________________________________________

\subsection{Minor bodies}

%-------------------------------------------------------------------

\subsubsection{Inner belt}\label{sec5.3.1:ResultsInnerBelt}

\begin{figure}
\centering
{\includegraphics[width=.49\linewidth]{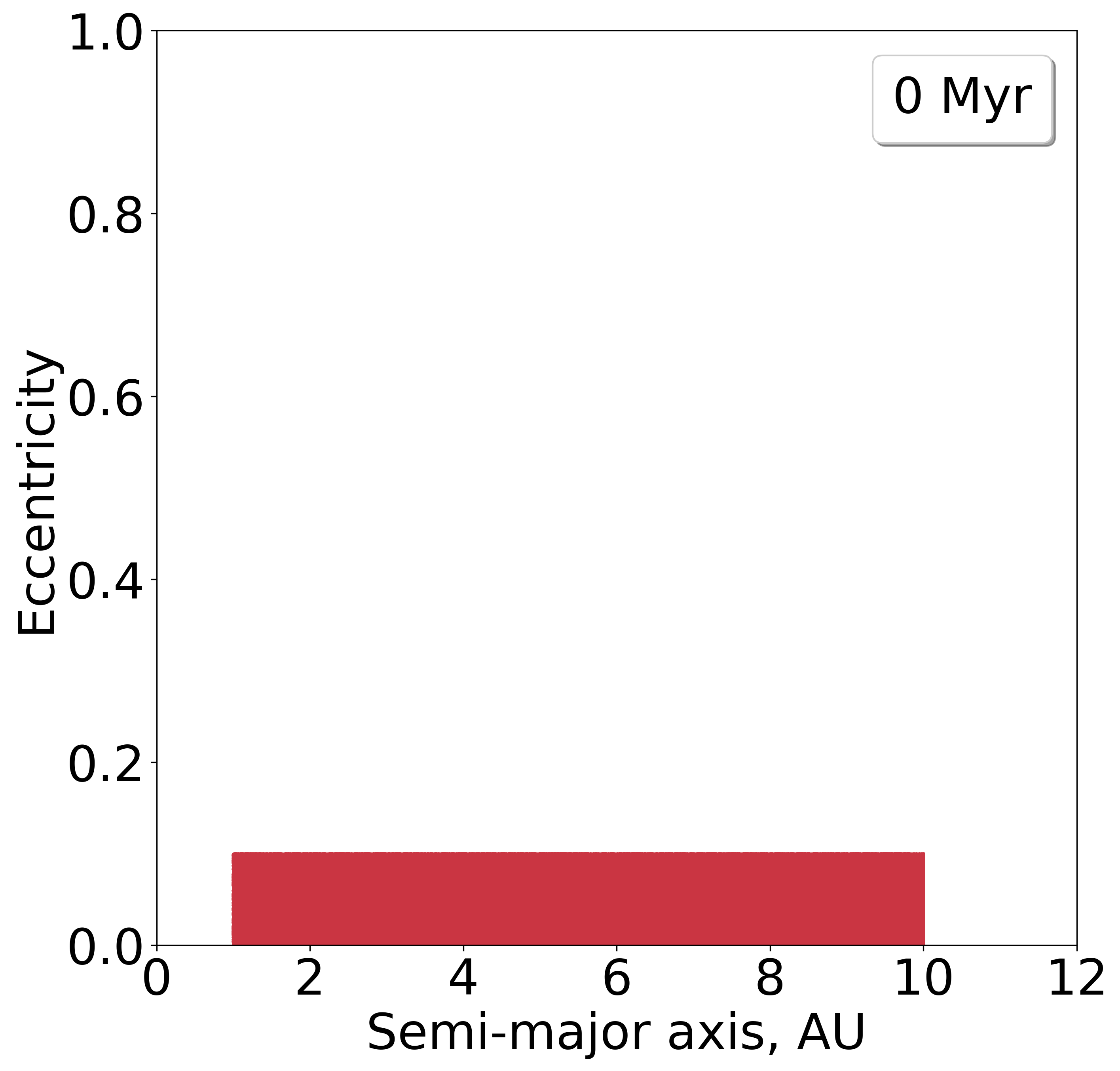}}
{\includegraphics[width=.49\linewidth]{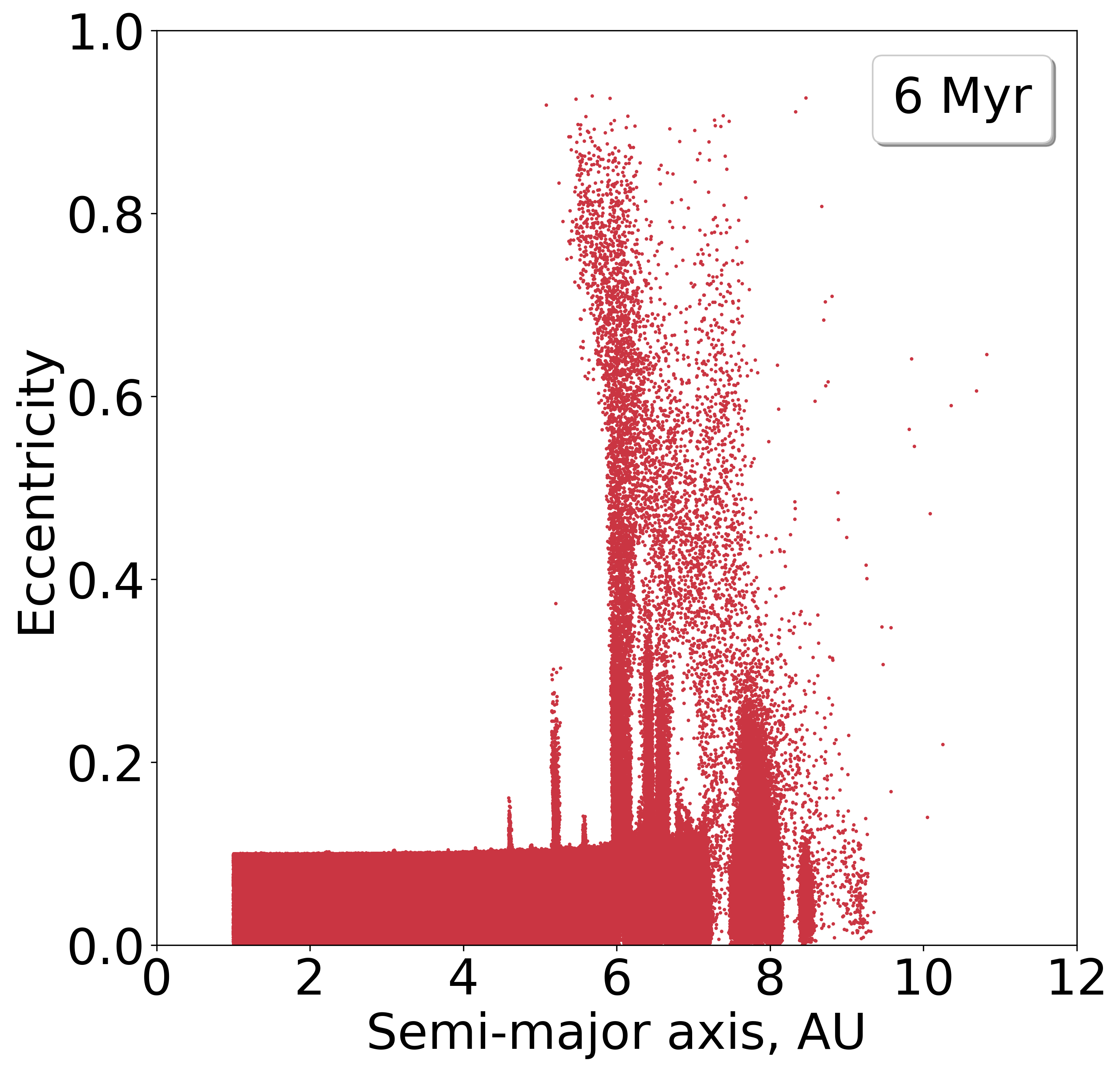}}
{\includegraphics[width=.49\linewidth]{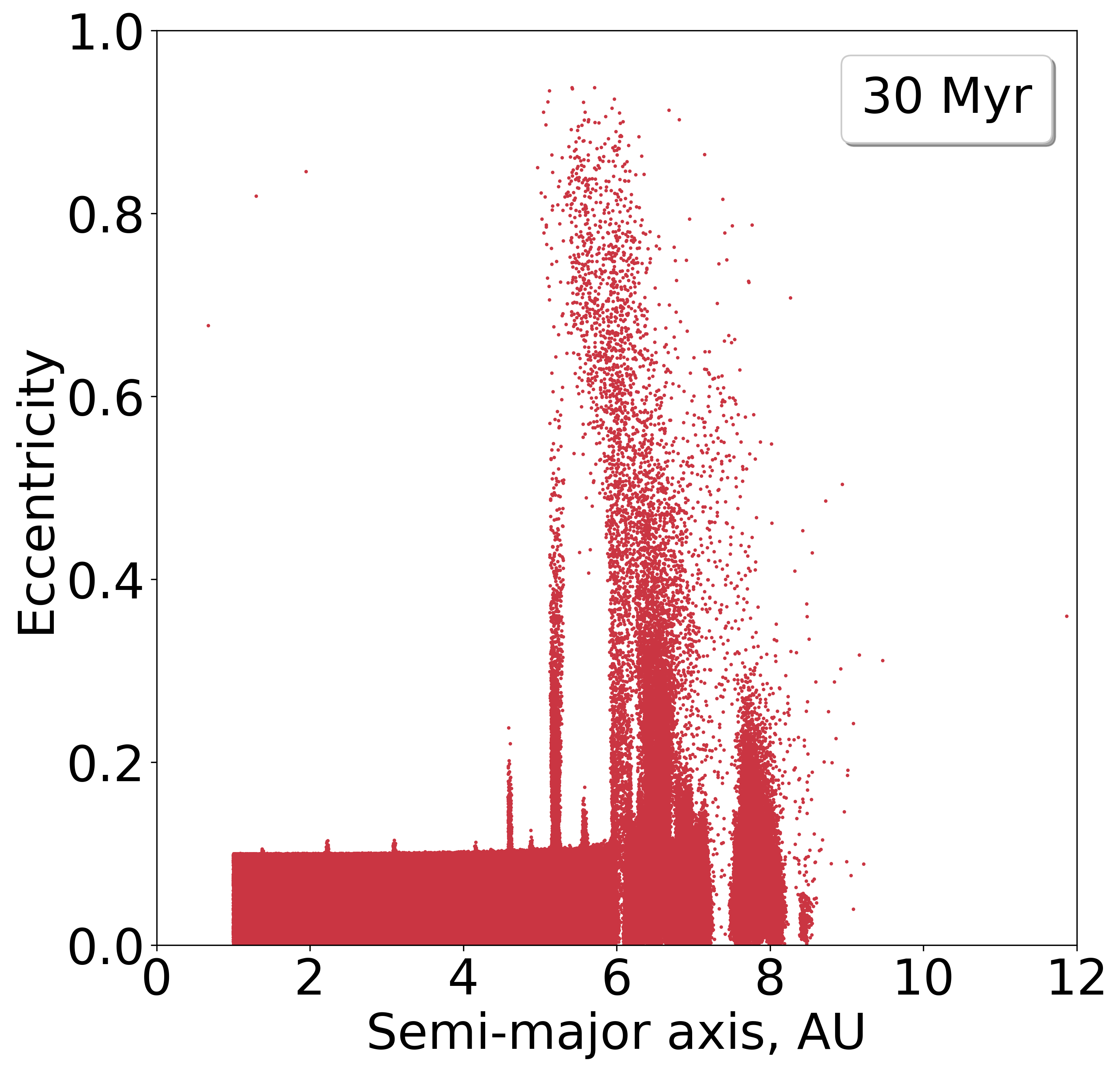}}
{\includegraphics[width=.49\linewidth]{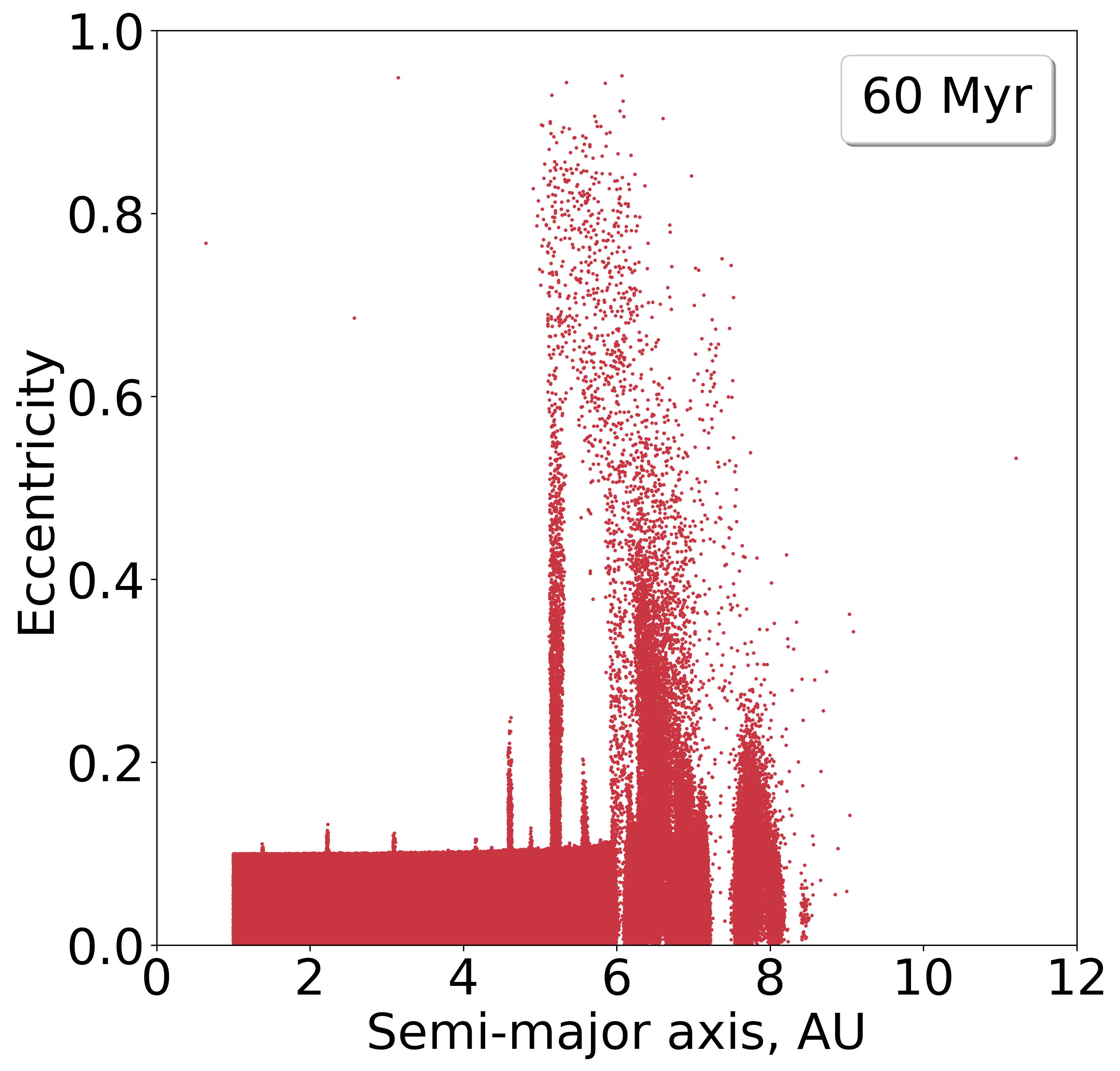}}
{\includegraphics[width=.49\linewidth]{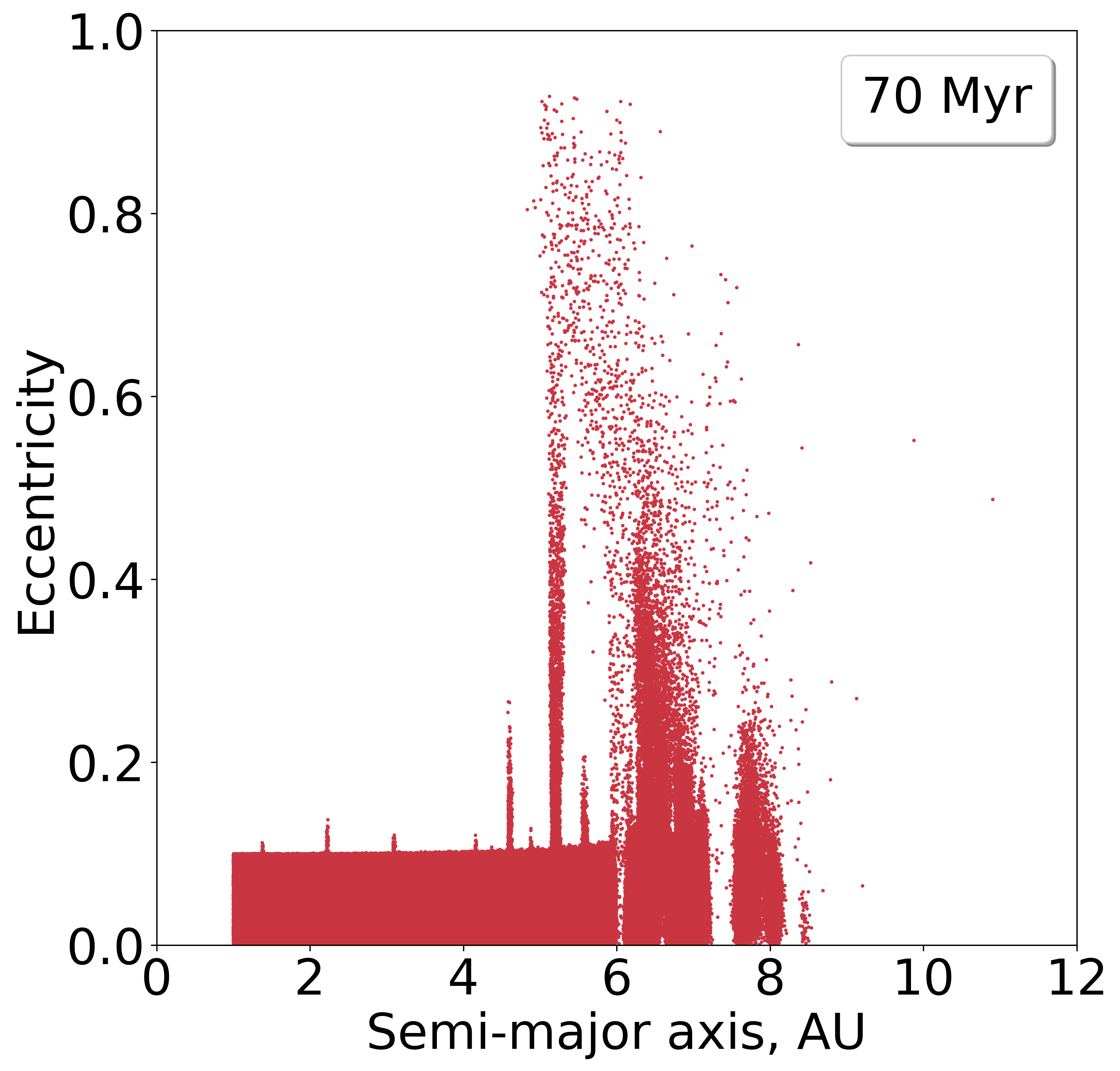}}
\caption{Time evolution of 650,000 test particles representing minor bodies in the inner belt for a period of 70~Myr. All four planets are located outside of the plot. The innermost planet HR~8799~e has a semi-major axis of 15.4~AU.}
\label{fig5.3.1:1}
\end{figure}

\begin{figure}
\centering
{\includegraphics[width=.49\linewidth]{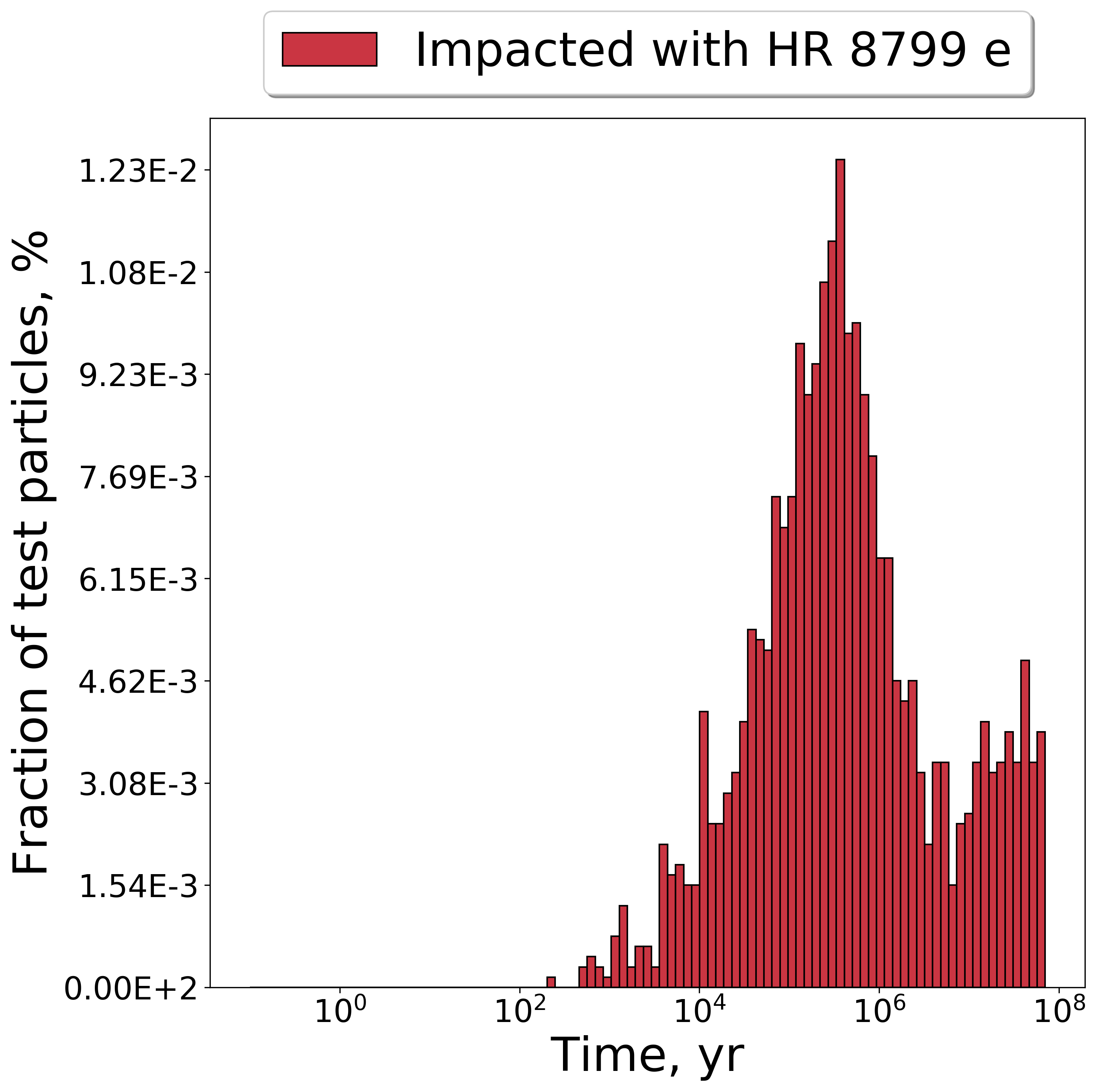}}
{\includegraphics[width=.49\linewidth]{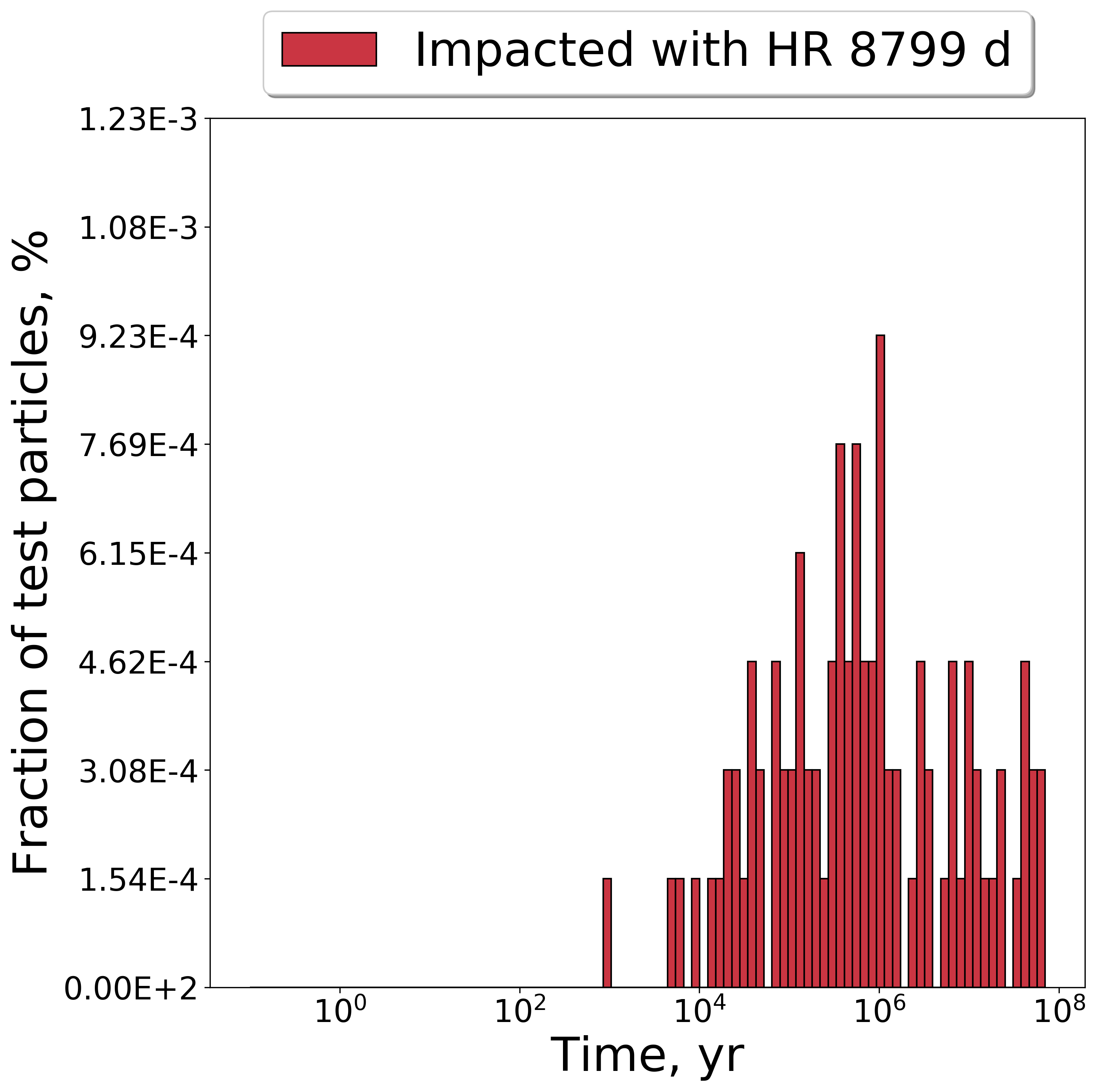}}
{\includegraphics[width=.49\linewidth]{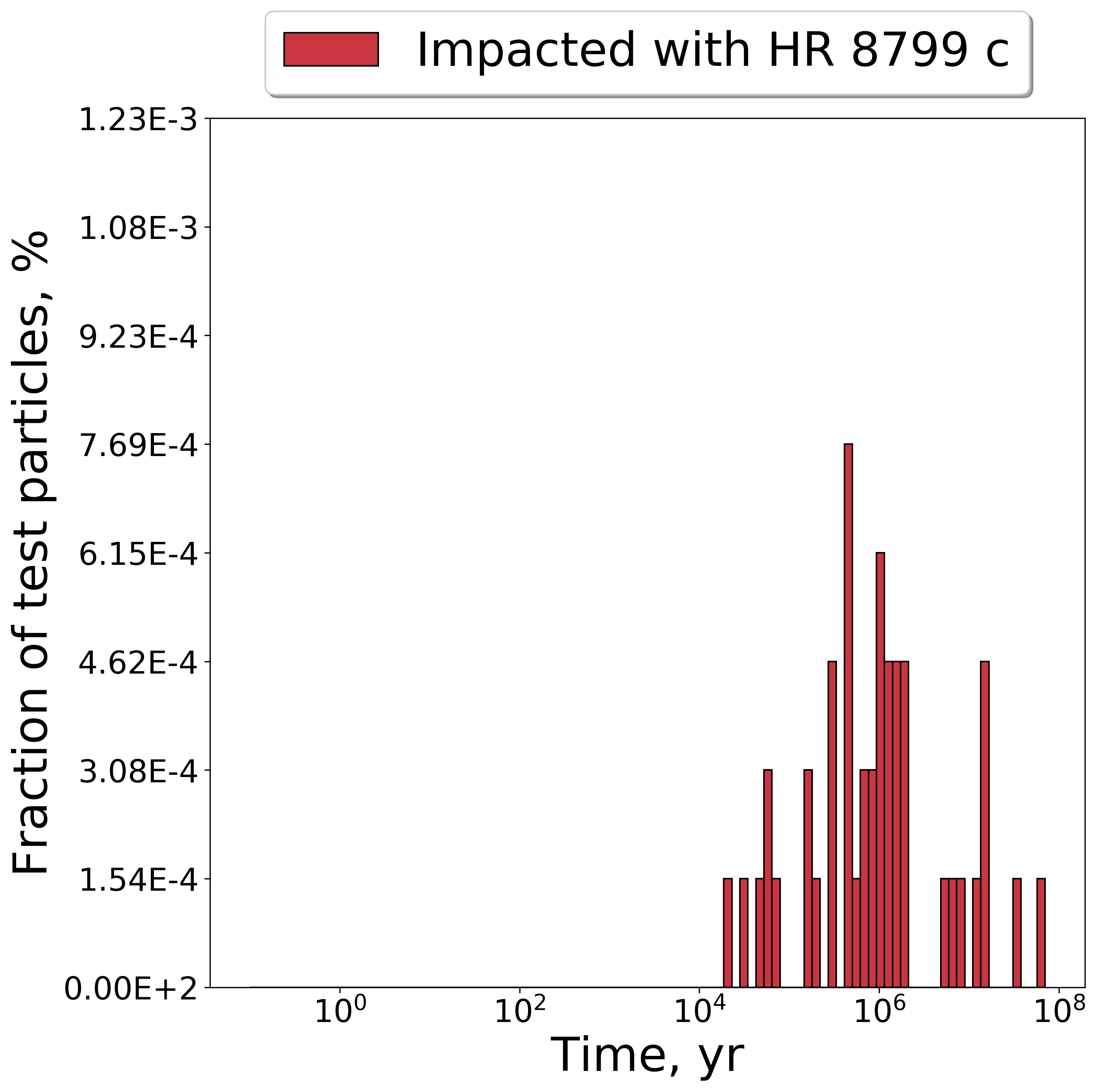}}
{\includegraphics[width=.49\linewidth]{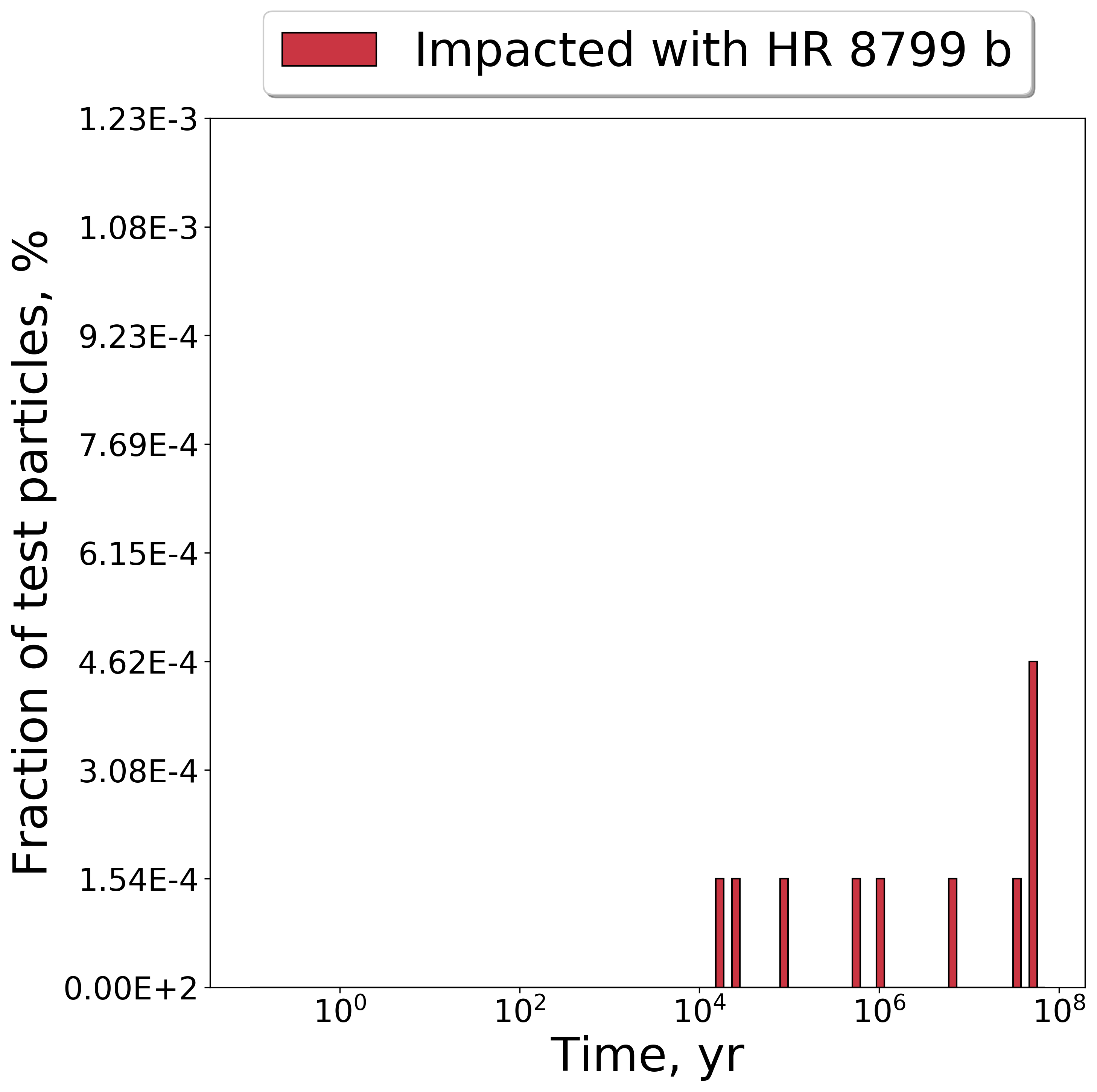}}
{\includegraphics[width=.49\linewidth]{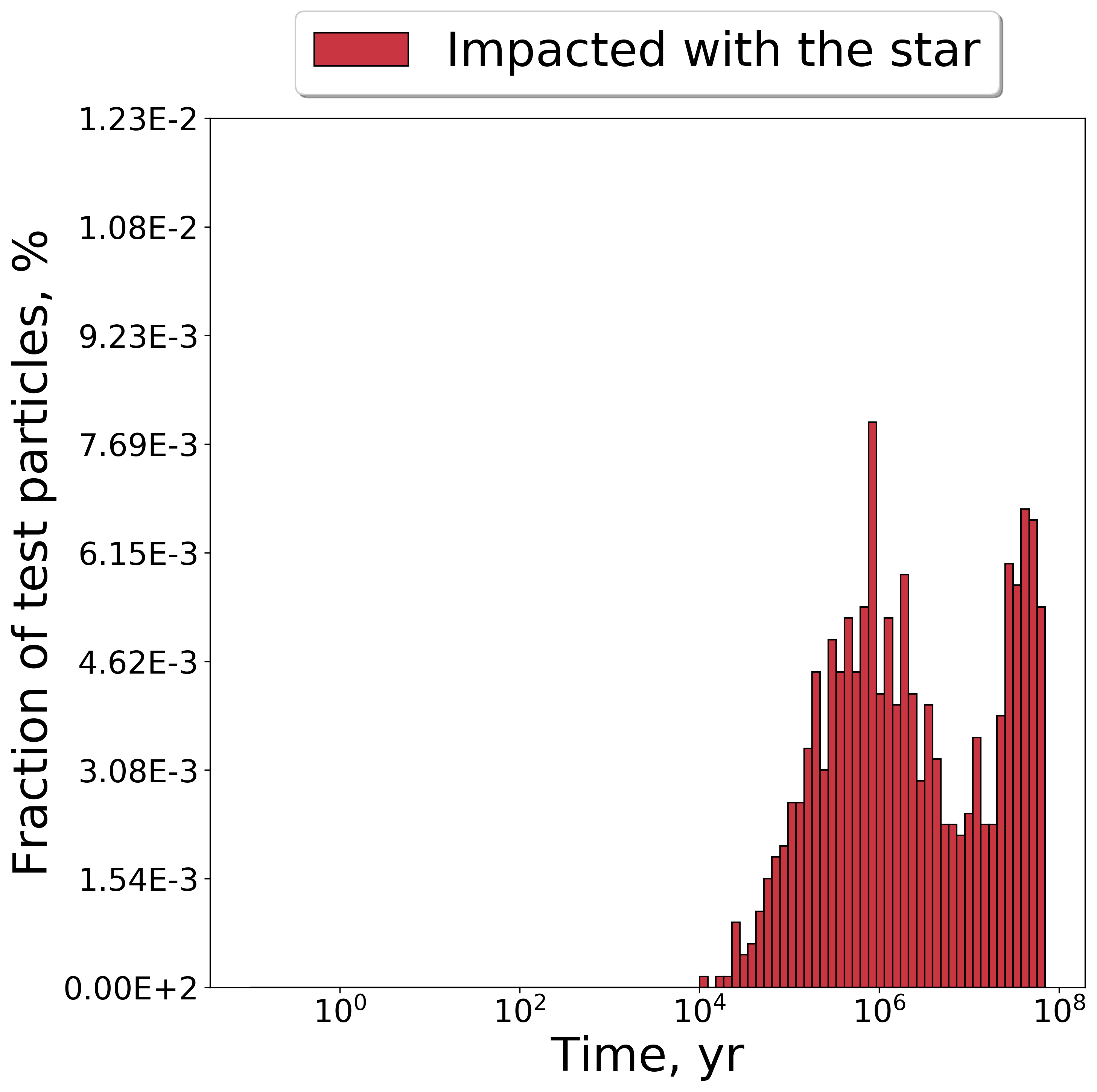}}
{\includegraphics[width=.49\linewidth]{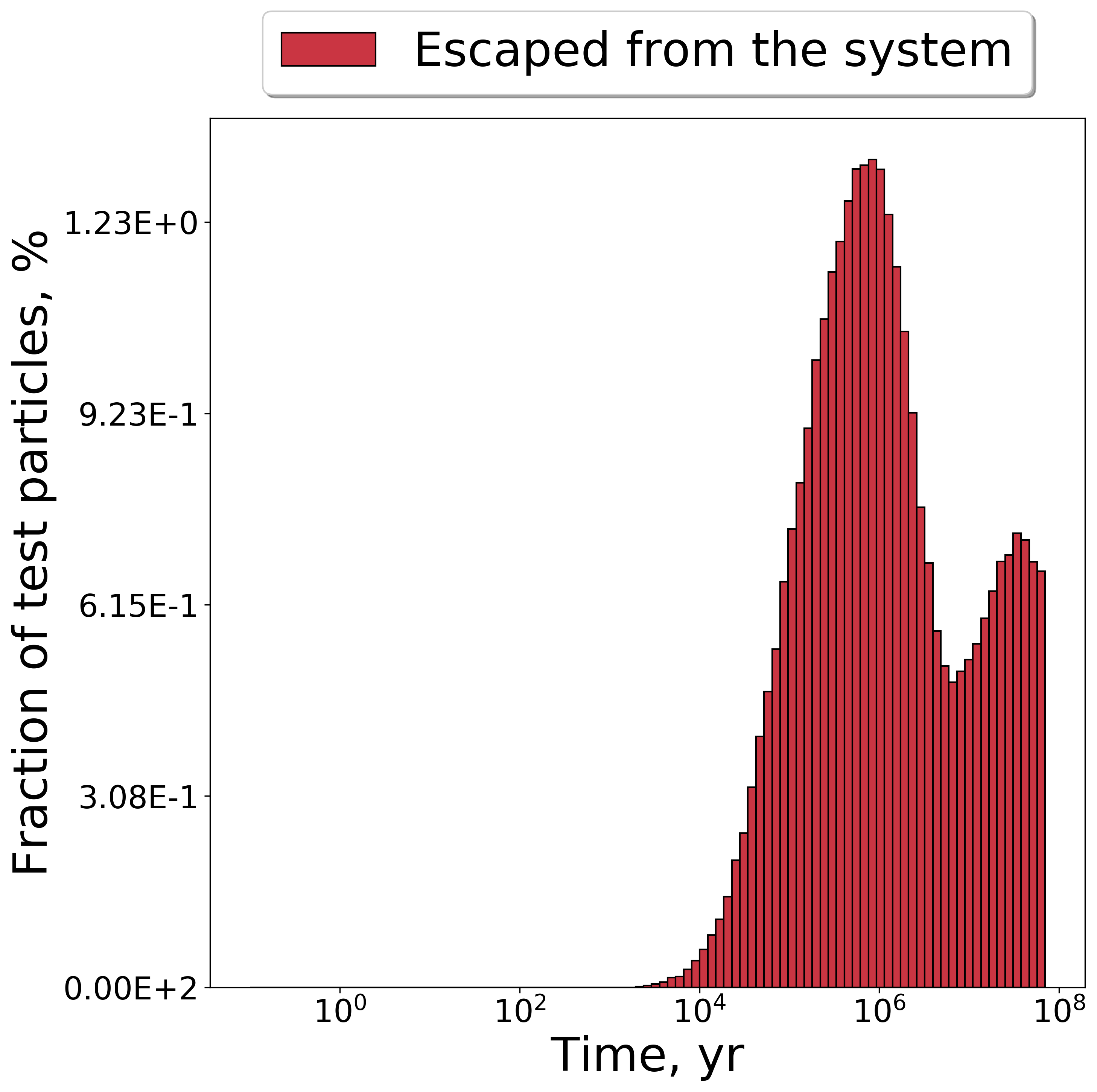}}
\caption{Fraction of test particles representing minor bodies originating from the inner belt that impacted with the planets, the star or exceeded a user-provided heliocentric distance, 1,000~AU, during 70~Myr of the simulations.}
\label{fig5.3.1:2}
\end{figure}

\begin{figure}
\centering
{\includegraphics[width=.49\linewidth]{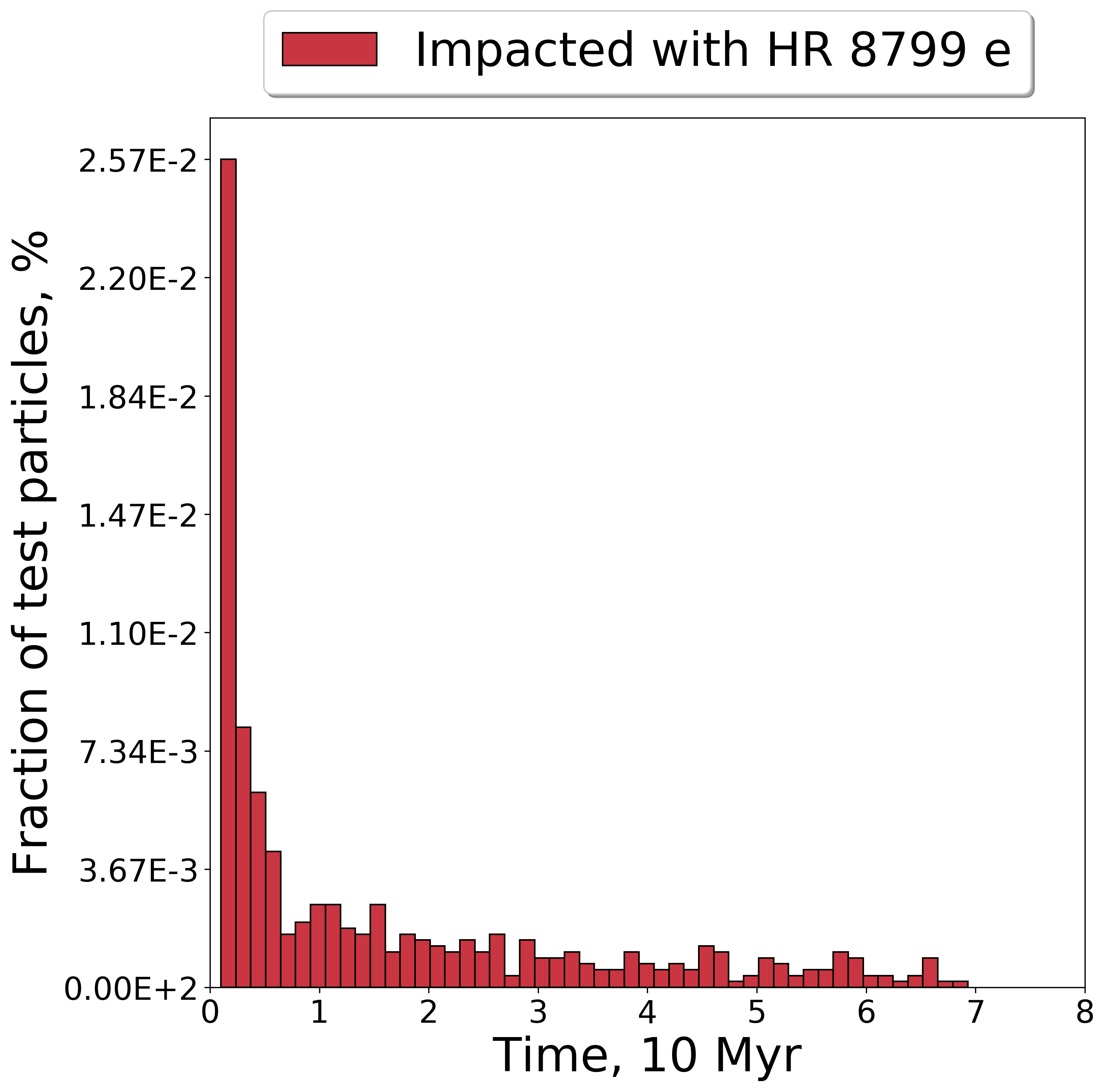}}
{\includegraphics[width=.49\linewidth]{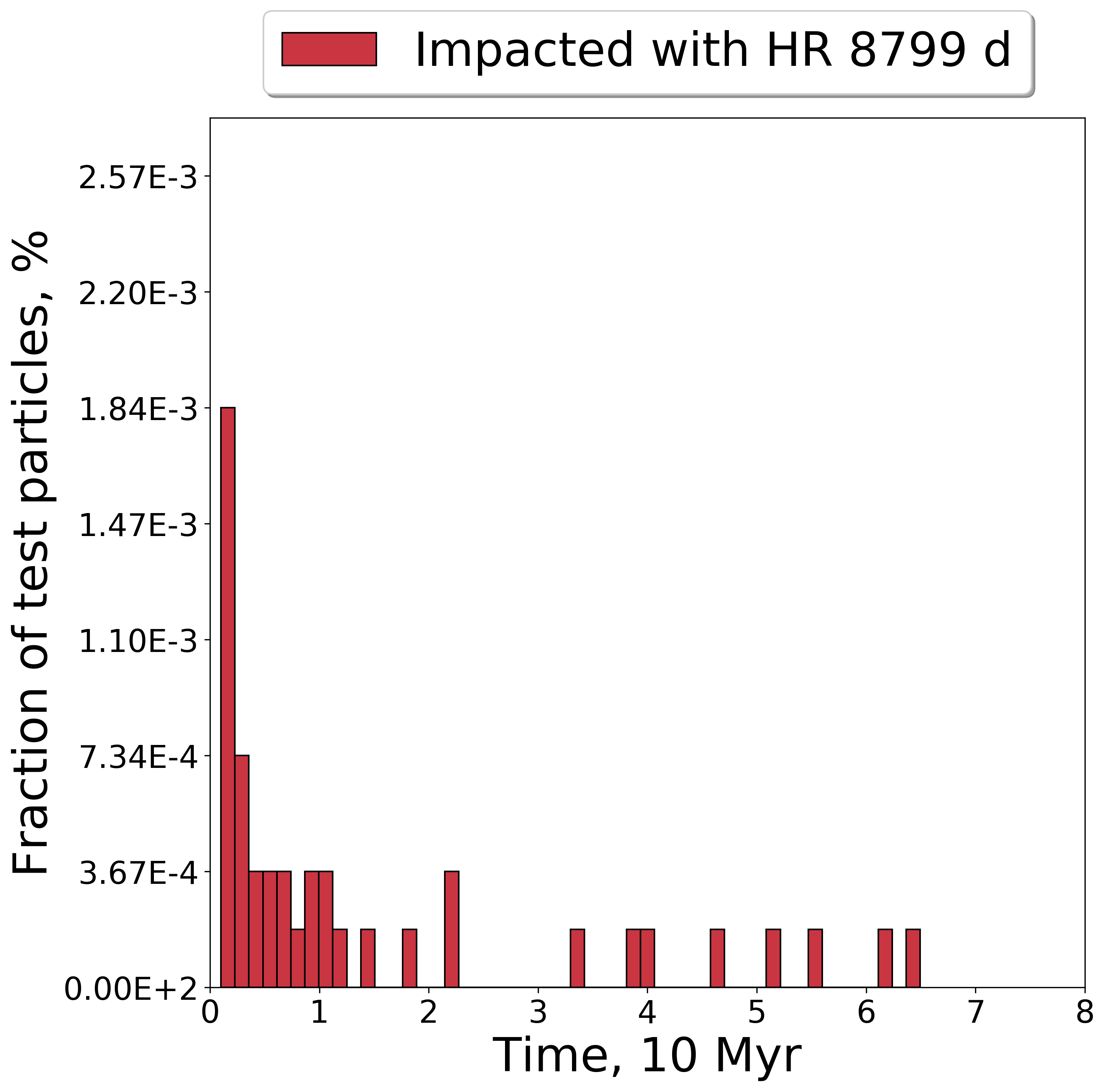}}
{\includegraphics[width=.49\linewidth]{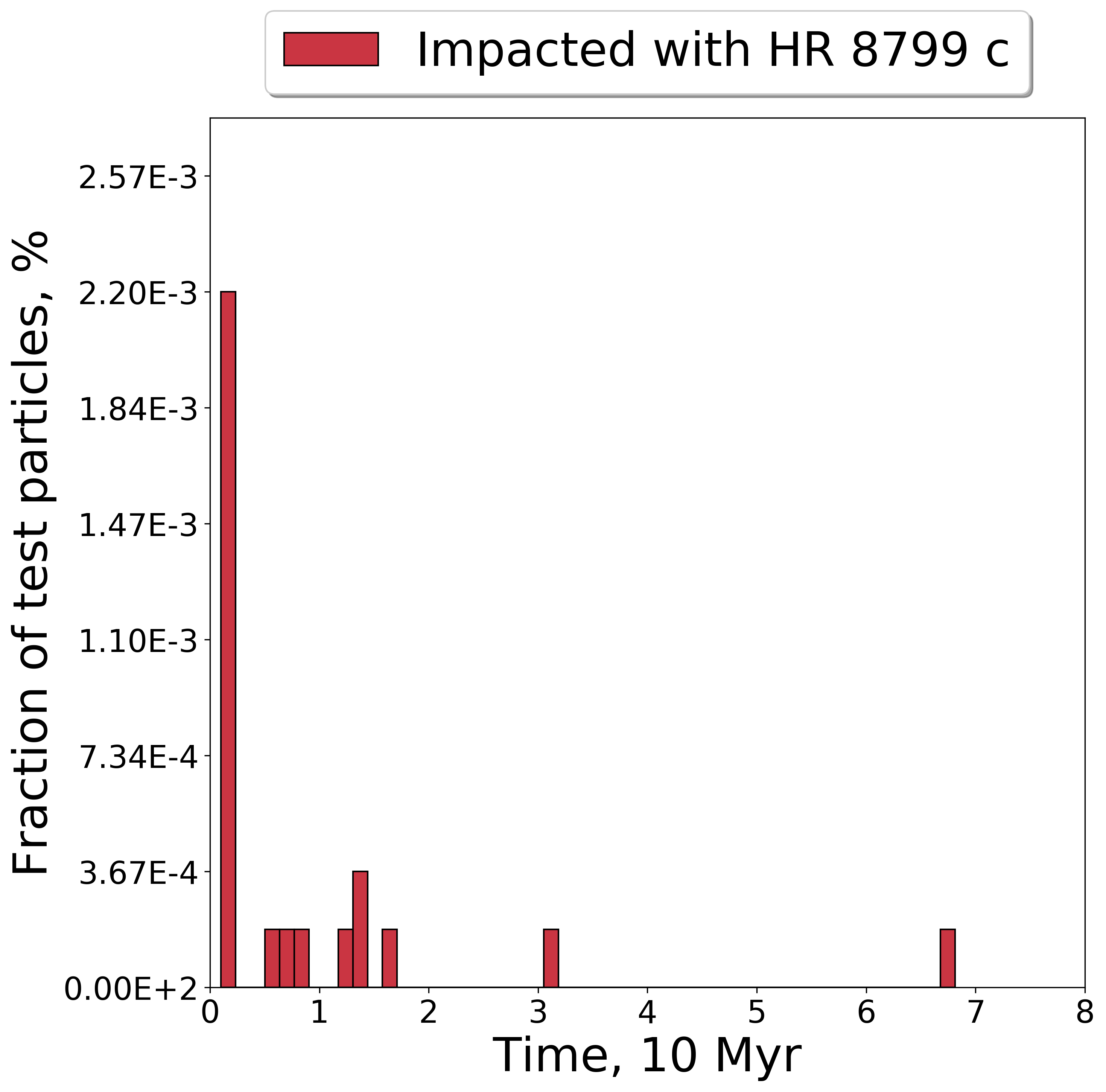}}
{\includegraphics[width=.49\linewidth]{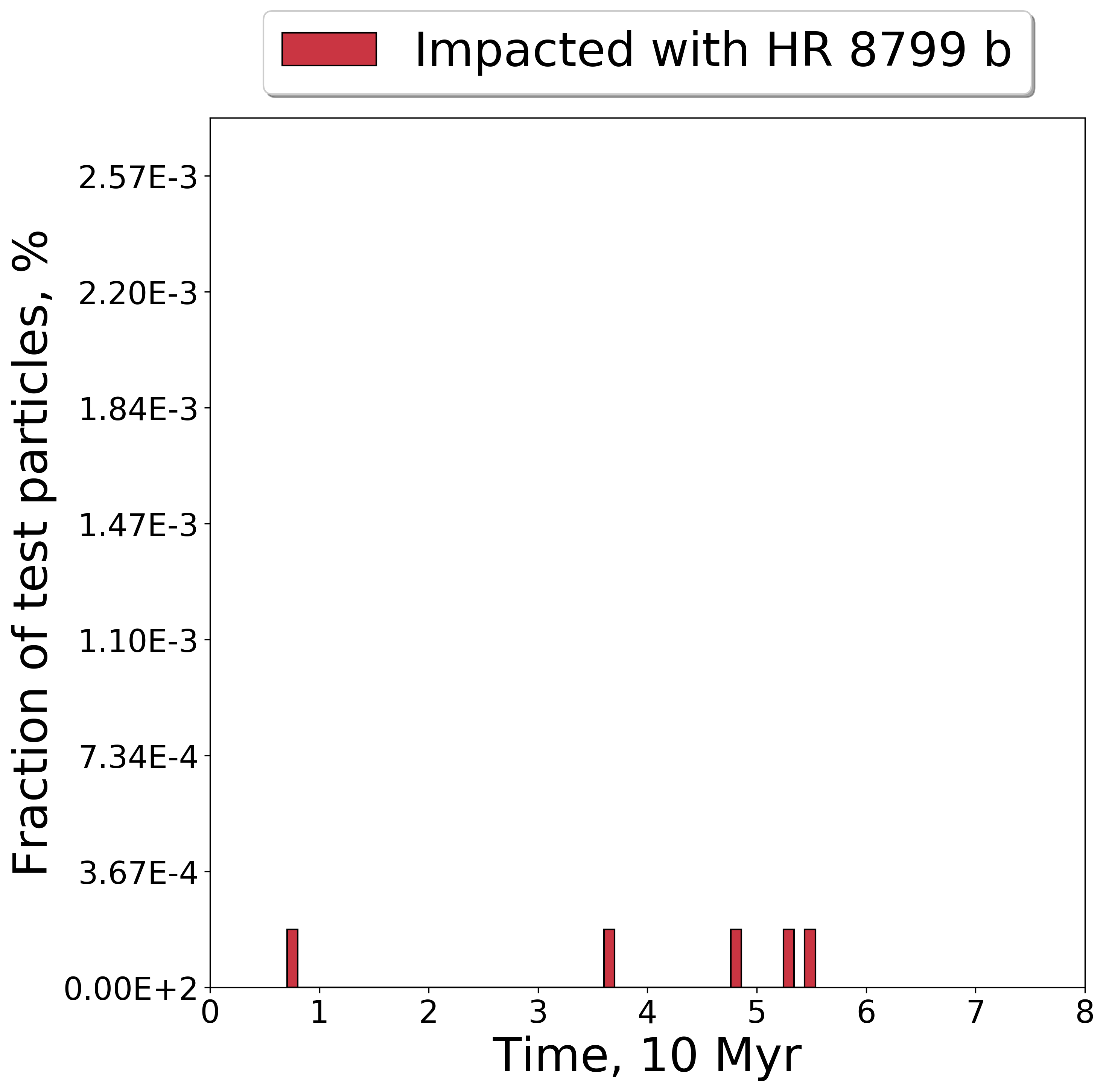}}
{\includegraphics[width=.49\linewidth]{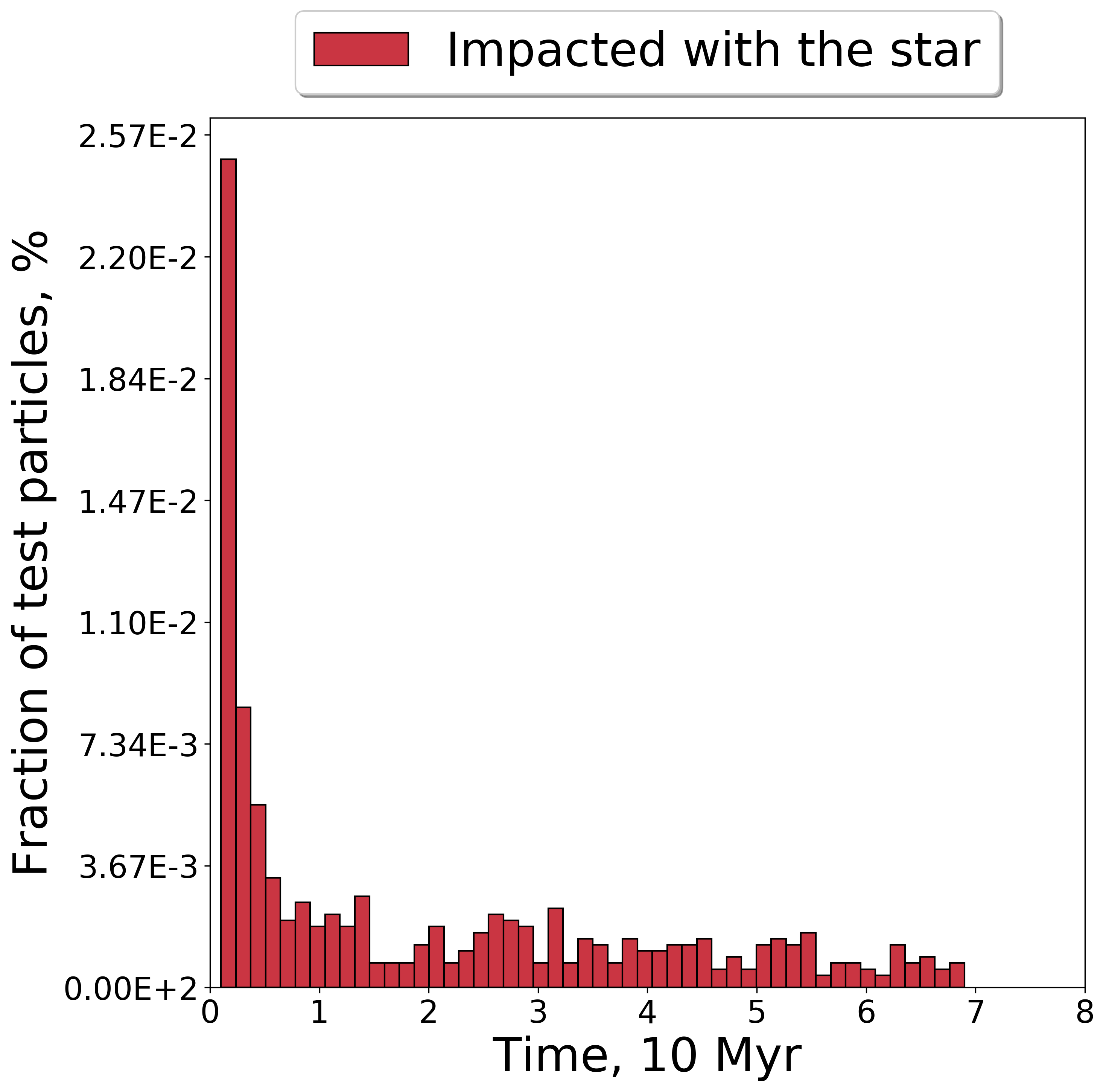}}
{\includegraphics[width=.49\linewidth]{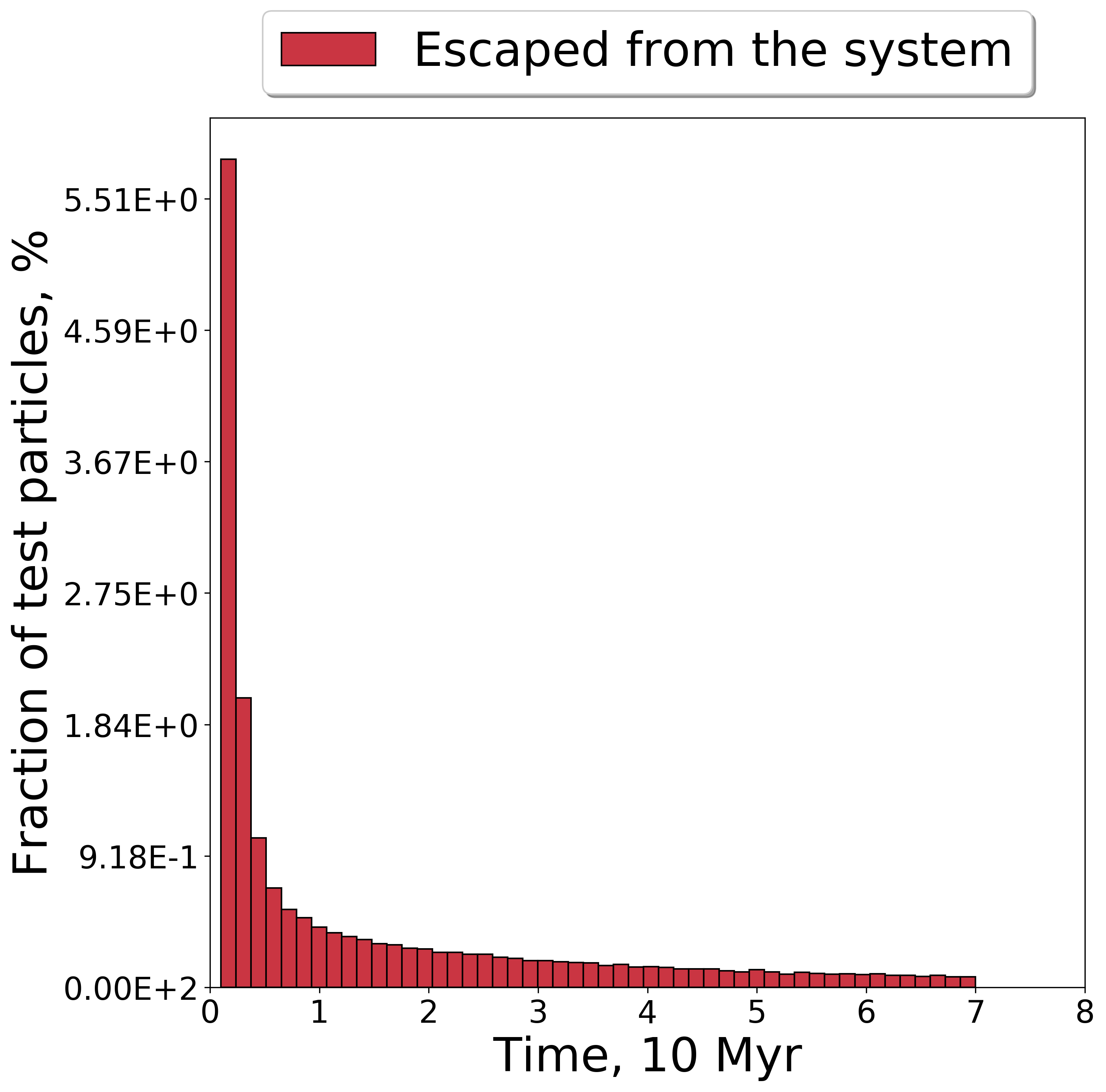}}
\caption{Same as Fig.~\ref{fig5.3.1:2}, but only after the first 1~Myr of the simulations, when the system reached steady state.}
\label{fig5.3.1:3}
\end{figure}

The snapshot at 6~Myr shows that around that time the inner belt develops structure due to the interaction between the test particles and the planets, see Fig.~\ref{fig5.3.1:1}. The interaction with the innermost planet HR~8799~e shapes the belt in a similar way Jupiter shapes our own Main Asteroid Belt. There are broad gaps at the locations of the mean motion resonances with the giant planets and there are many dynamically excited objects with large eccentricities and inclinations. At the end of the simulation, after 70~Myr, 0.143\% of test particles were discarded due to collisions with the star and 31.224\% of test particles were found to be ejected from the system. As shown in Table~\ref{tbl5.3.1:1}, 0.248\% of test particles collided with the innermost planet HR~8799~e, 0.014\% with HR~8799~d, 0.007\% with HR~8799~c, and 0.002\% with HR~8799~b. As seen from Fig.~\ref{fig5.3.1:2}, the number of  impacts peaks prominently in the first 1~Myr. We attribute this to test particles that were initialised on highly unstable orbits. After the first $\sim 1$~Myr, impacts continue at a roughly continuous rate, see Fig.~\ref{fig5.3.1:3}; this is referred to as steady state in the following. The steady state impact rates is what we are interested in for the purposes of this project.

The initial conditions of the inner belt simulations followed the example of \citet{Contro2016}. As can be seen in our Figure~\ref{fig5.3.1:1} \citep[see Fig.~3 in][]{Contro2016}, we reproduce the resulting structure of the inner belt from their simulations. At the end of the simulations there are almost no particles left beyond 8~AU. At the location of the 3:1 mean motion resonance with the innermost planet, at $\sim7.4$~AU, a broad gap is formed as well as a smaller gap at the location of the 4:1 mean motion resonance ($\sim6$~AU). Unlike \citeauthor{Contro2016} we resolve and analyse impacts between planets and test particles, see \S\ref{sec5.4:VolatileDeliveryRates}.

\begin{table*}
\caption{Percentage of test particles representing minor bodies that were discarded from the simulations of both belts due to the minor body collisions with the star and four planets, and due to exceeding the maximum distance from the star with respect to the initial number of test particles. In brackets we state the absolute number of test particles. We note that the initial number of test particles for 0-70\,Myr is larger than for 1-70\,Myr.}
\label{tbl5.3.1:1}
\centering
\begin{tabular}{r|r|r|rrrrr}
\hline
\multicolumn{1}{c|}{\small simulation time} & \multicolumn{1}{c|}{\small belt} & \multicolumn{1}{c|}{\small ejection} & \multicolumn{5}{c}{\small collision} \\
\hline 
& &  & {\small star} & {\small e} & {\small d} & {\small c} & {\small b} \\
\hline
{\small0-70~Myr} & {\small inner} & {\small 31\% (202,958)} & {\small 0.14\% (928)} & {\small 0.25\% (1,614)} & {\small 0.014\% (92)} & {\small 0.0068\% (44)} & {\small 0.0015\% (10)} \\
& {\small outer} & {\small 9.5\% (137,143)} & {\small 0.014\% (208)} & {\small 0.0037\% (54)} & {\small 0.0049\% (72)} & {\small 0.0077\% (112)} & {\small 0.041\% (595)} \\
\hline 
{\small1-70~Myr} & {\small inner} & {\small 18\% (99,202)} & {\small 0.1\% (545)} & {\small 0.091\% (494)} & {\small 0.0069\% (38)} & {\small 0.0039\% (21)} & {\small 0.00092\% (5)} \\
& {\small outer} & {\small 2.6\% (35,205)} & {\small 0.0013\% (17)} & {\small 0.00029\% (4)} & {\small 0.00069\% (9)} & {\small 0.00045\% (6)} & {\small 0.0013\% (18)} \\
\hline
\end{tabular}
\end{table*}

%__________________________________________________________________

\subsubsection{Outer belt}\label{sec5.3.2:ResultsOuterBelt}

\begin{figure}
\centering
{\includegraphics[width=.49\linewidth]{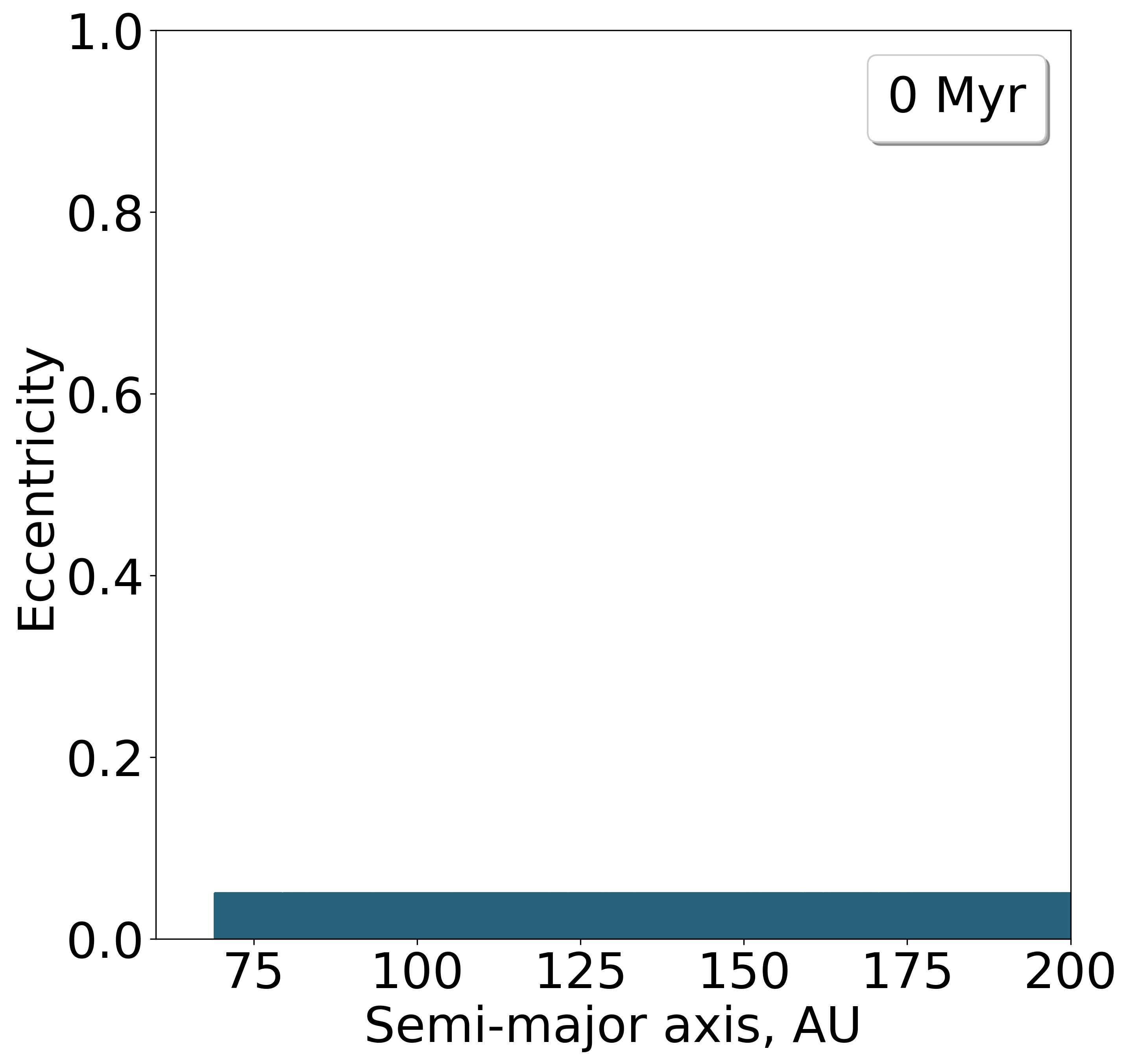}}
{\includegraphics[width=.49\linewidth]{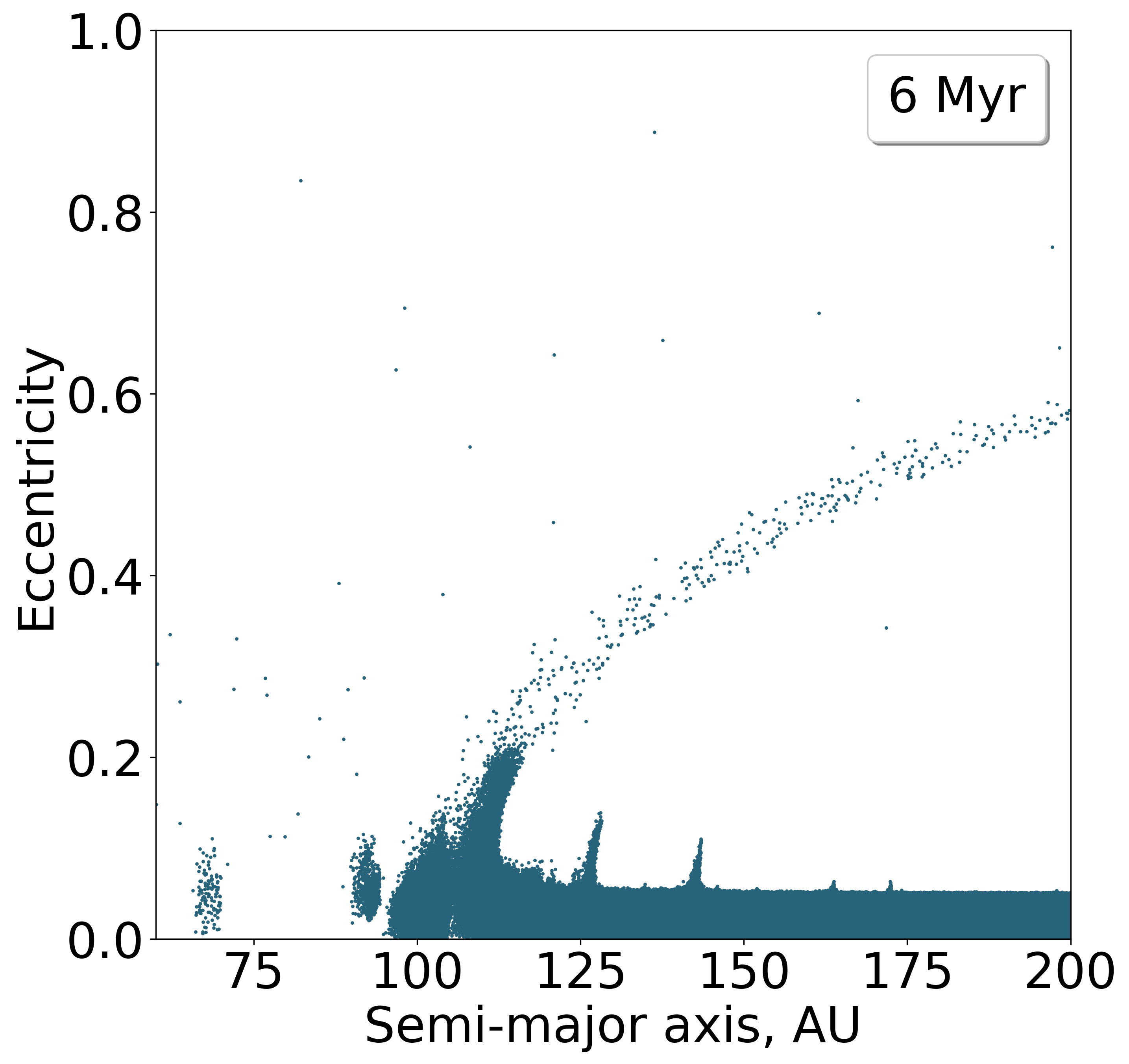}}
{\includegraphics[width=.49\linewidth]{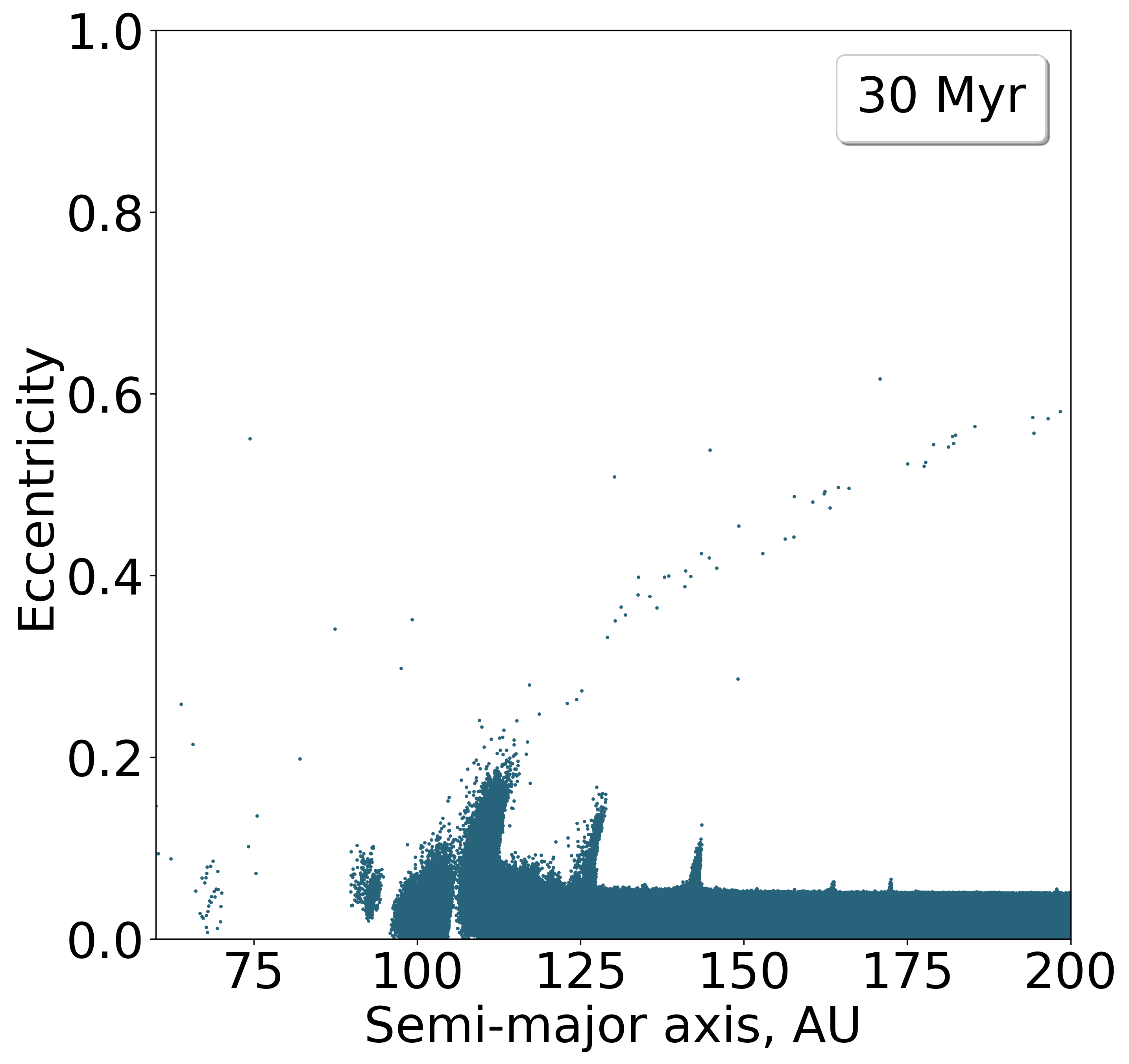}}
{\includegraphics[width=.49\linewidth]{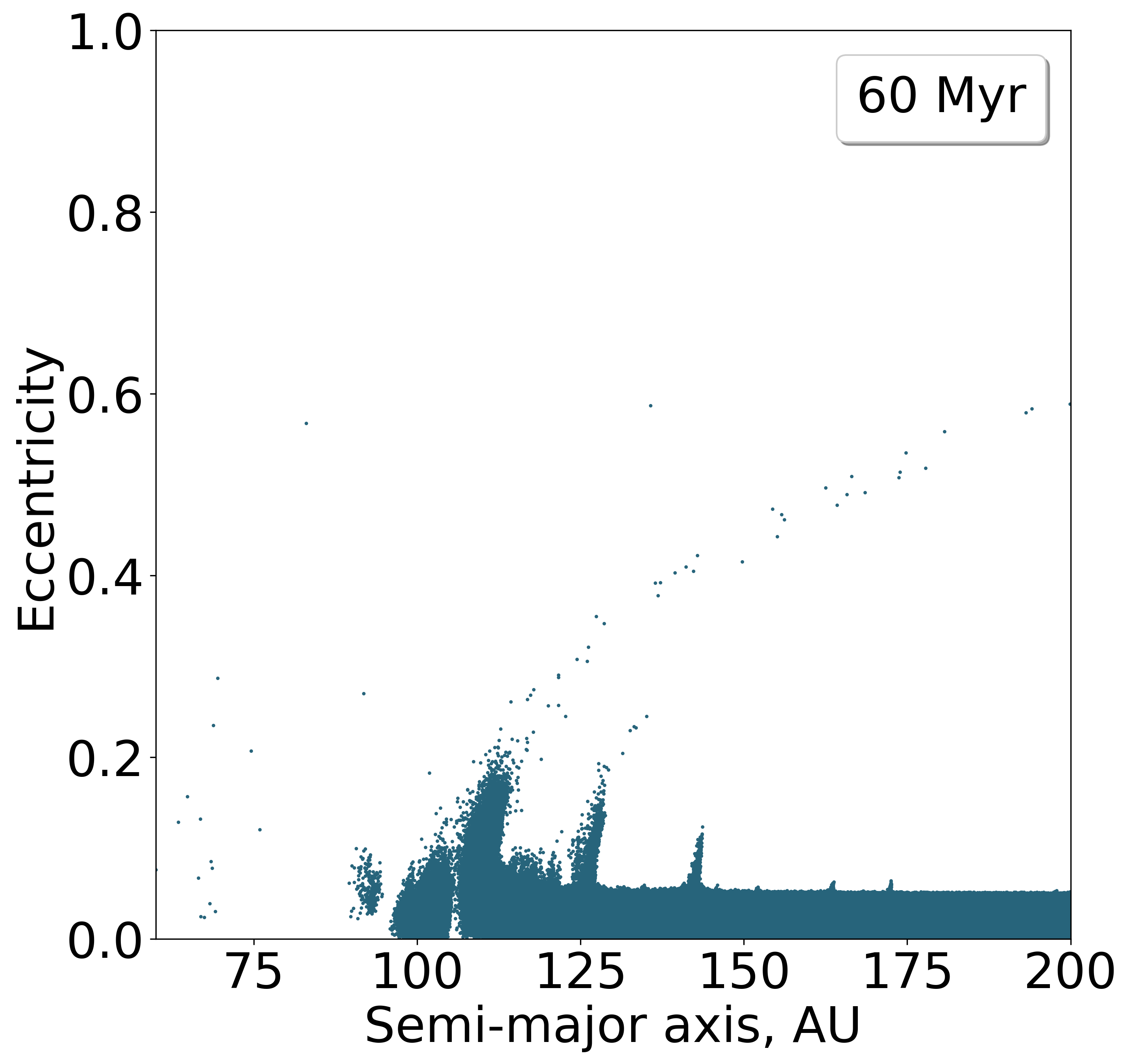}}
{\includegraphics[width=.49\linewidth]{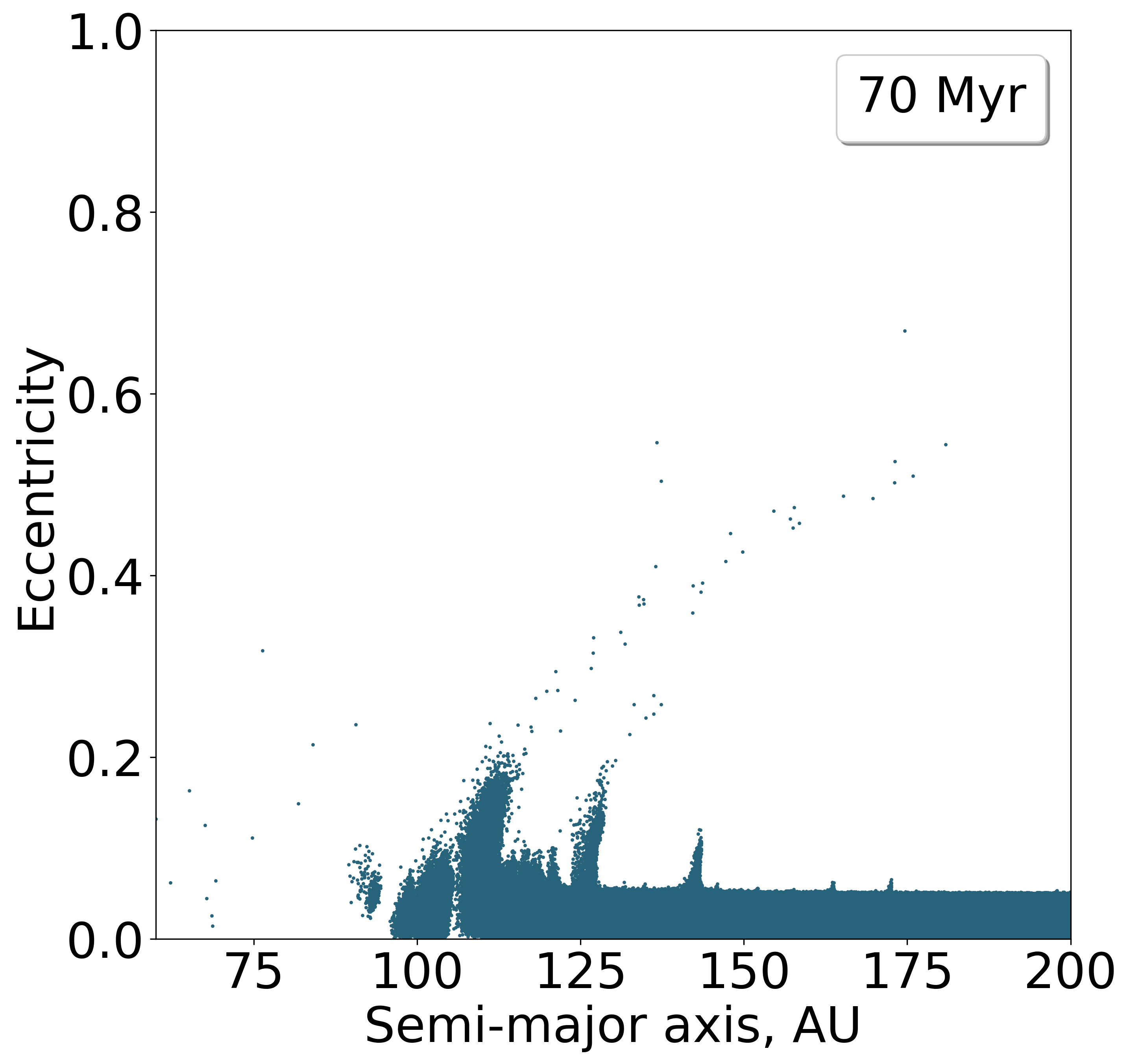}}
\caption{Same as Fig.~\ref{fig5.3.1:1}, but for the outer belt. 70~Myr evolution of 1,450,000 test particles representing minor bodies. The outermost planet HR~8799~b has a semi-major axis of 69.1~AU.}
\label{fig5.3.2:1}
\end{figure}

\begin{figure}
\centering
{\includegraphics[width=.49\linewidth]{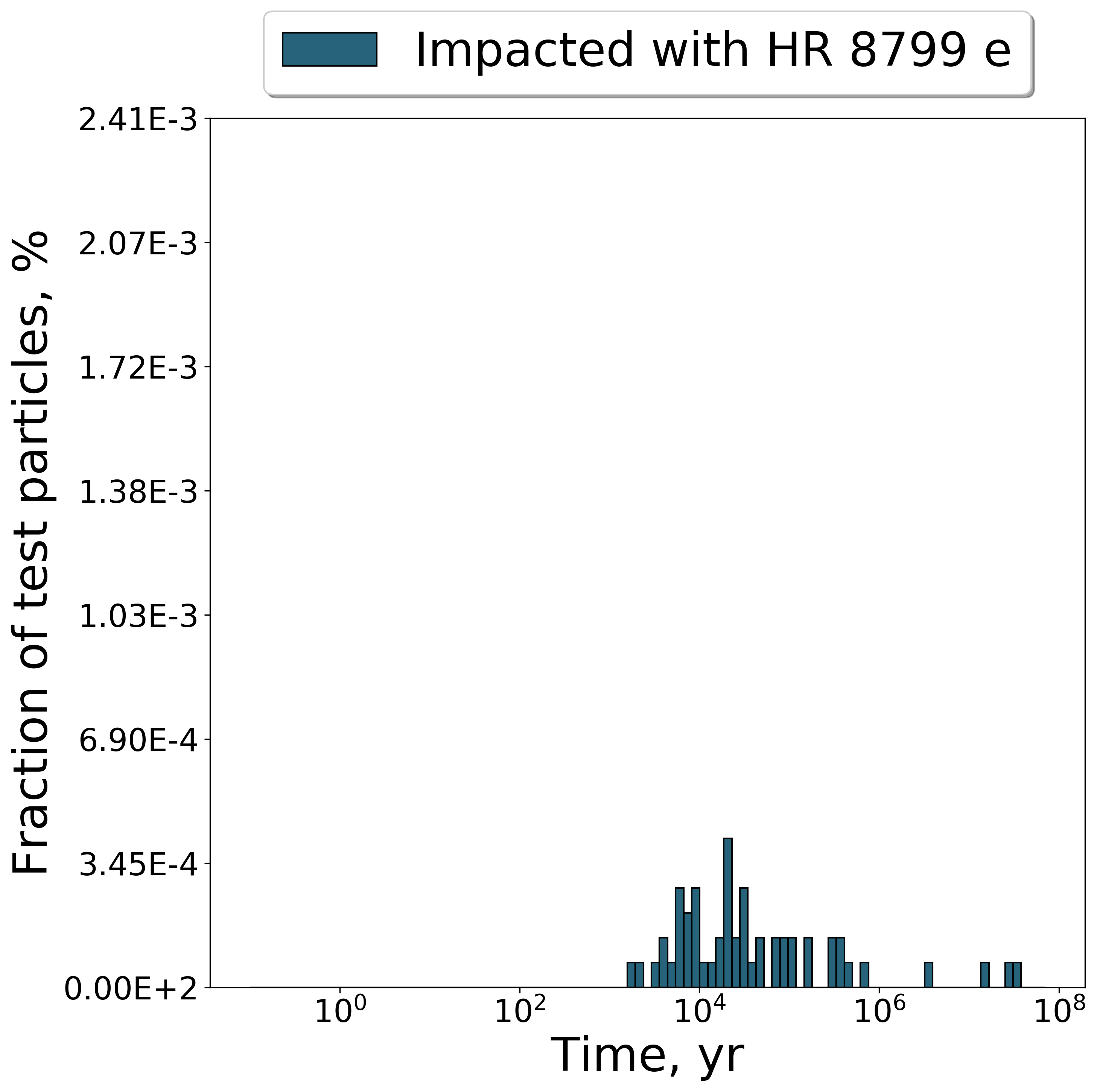}}
{\includegraphics[width=.49\linewidth]{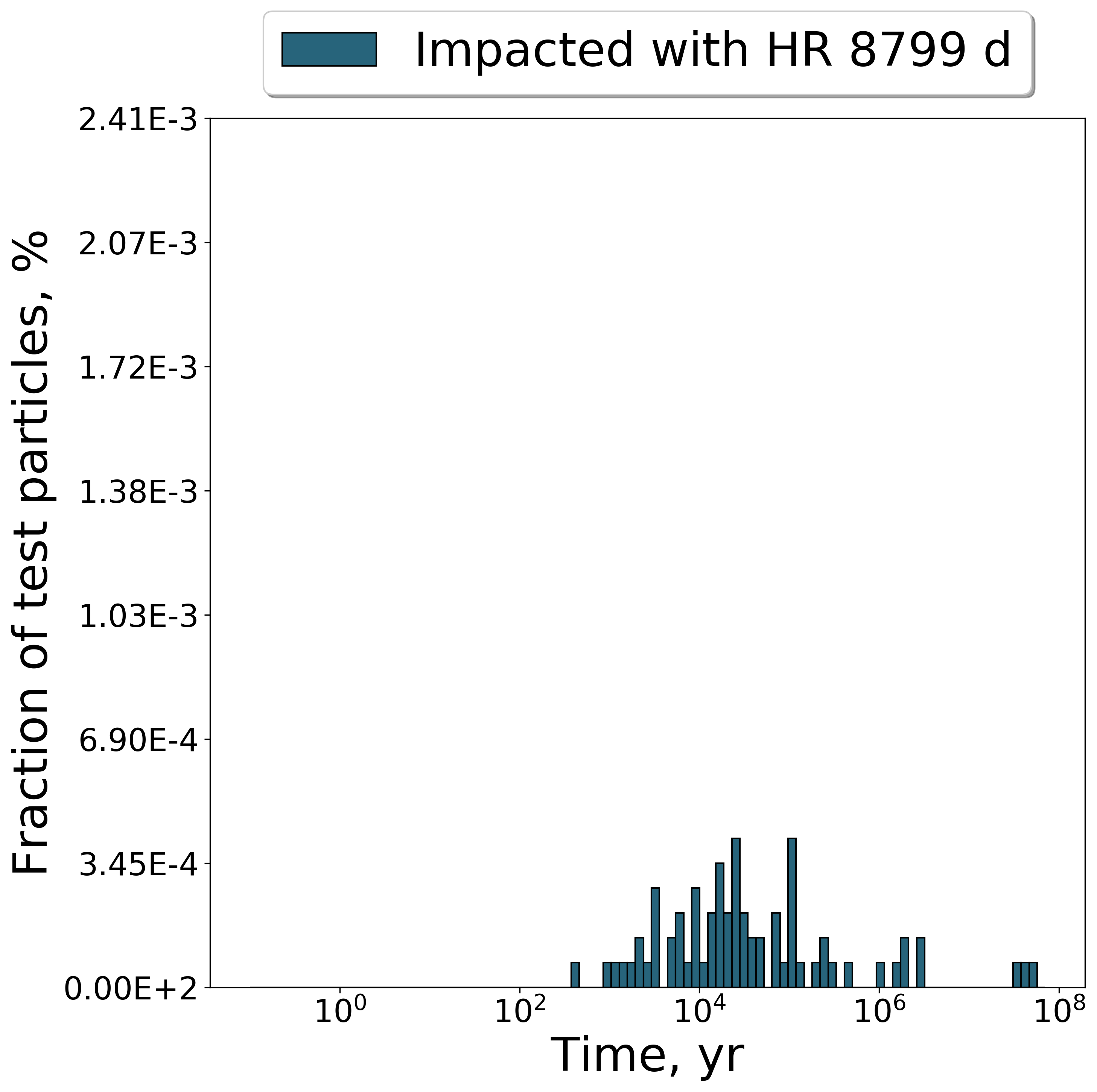}}
{\includegraphics[width=.49\linewidth]{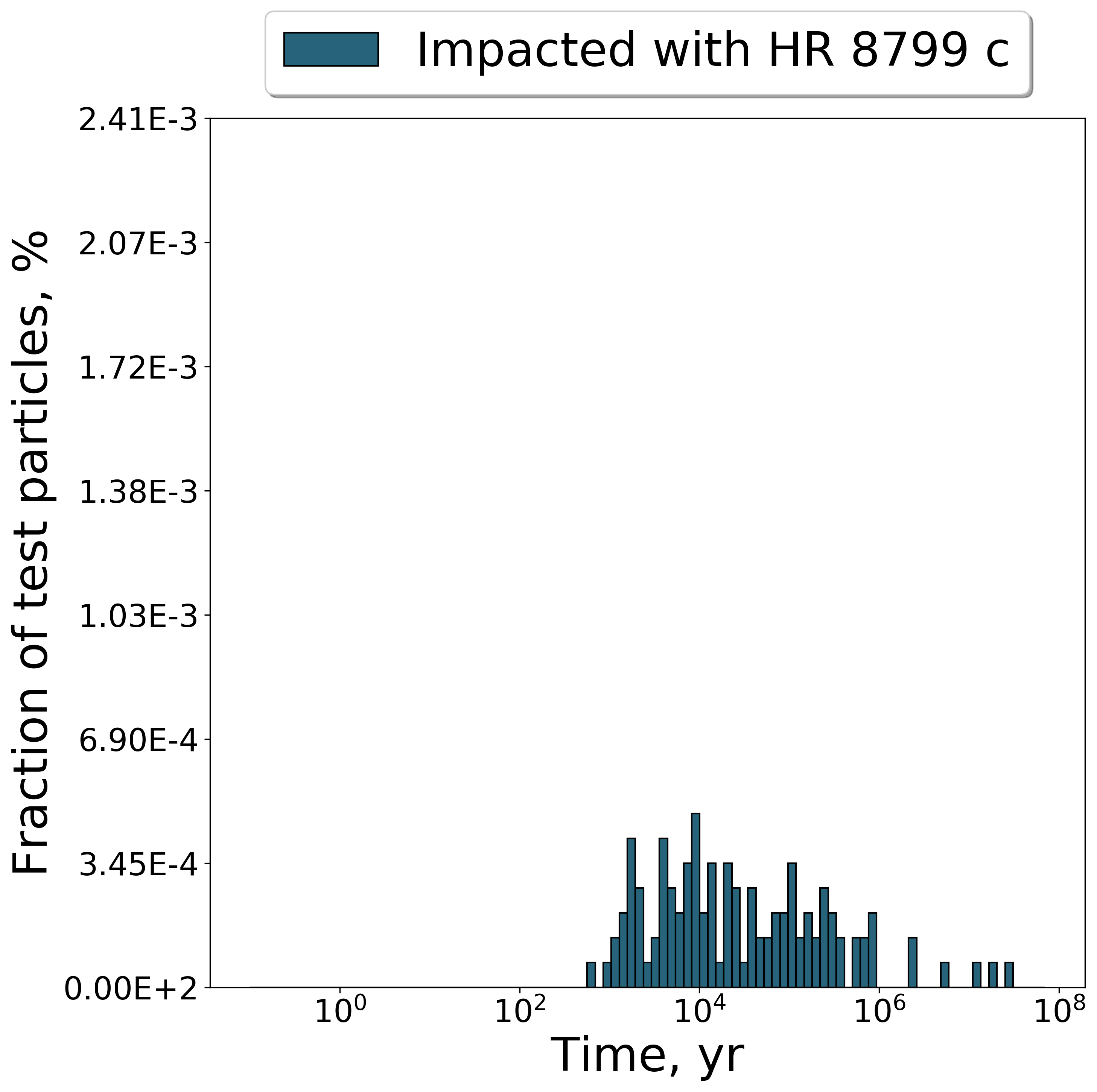}}
{\includegraphics[width=.49\linewidth]{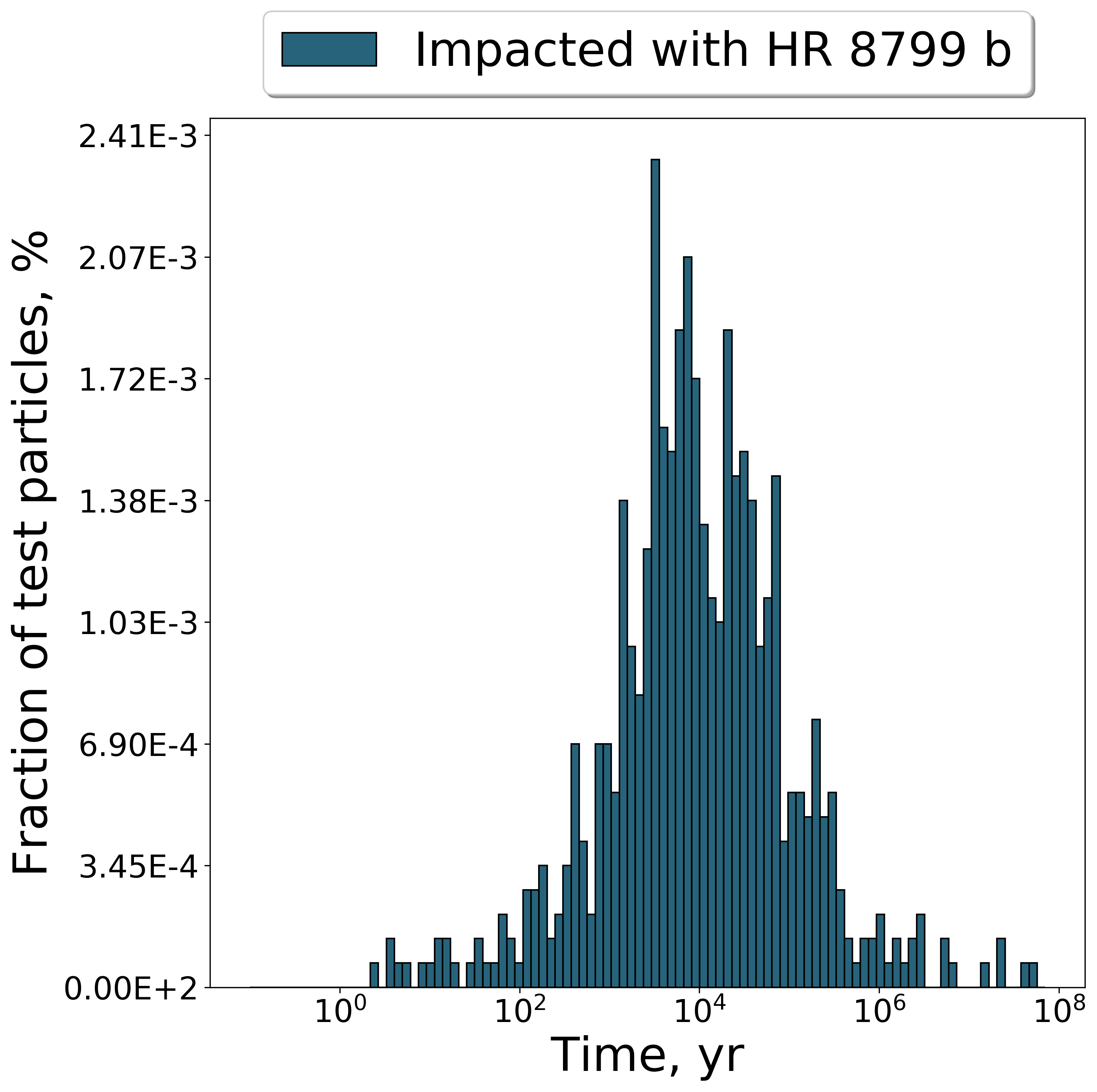}}
{\includegraphics[width=.49\linewidth]{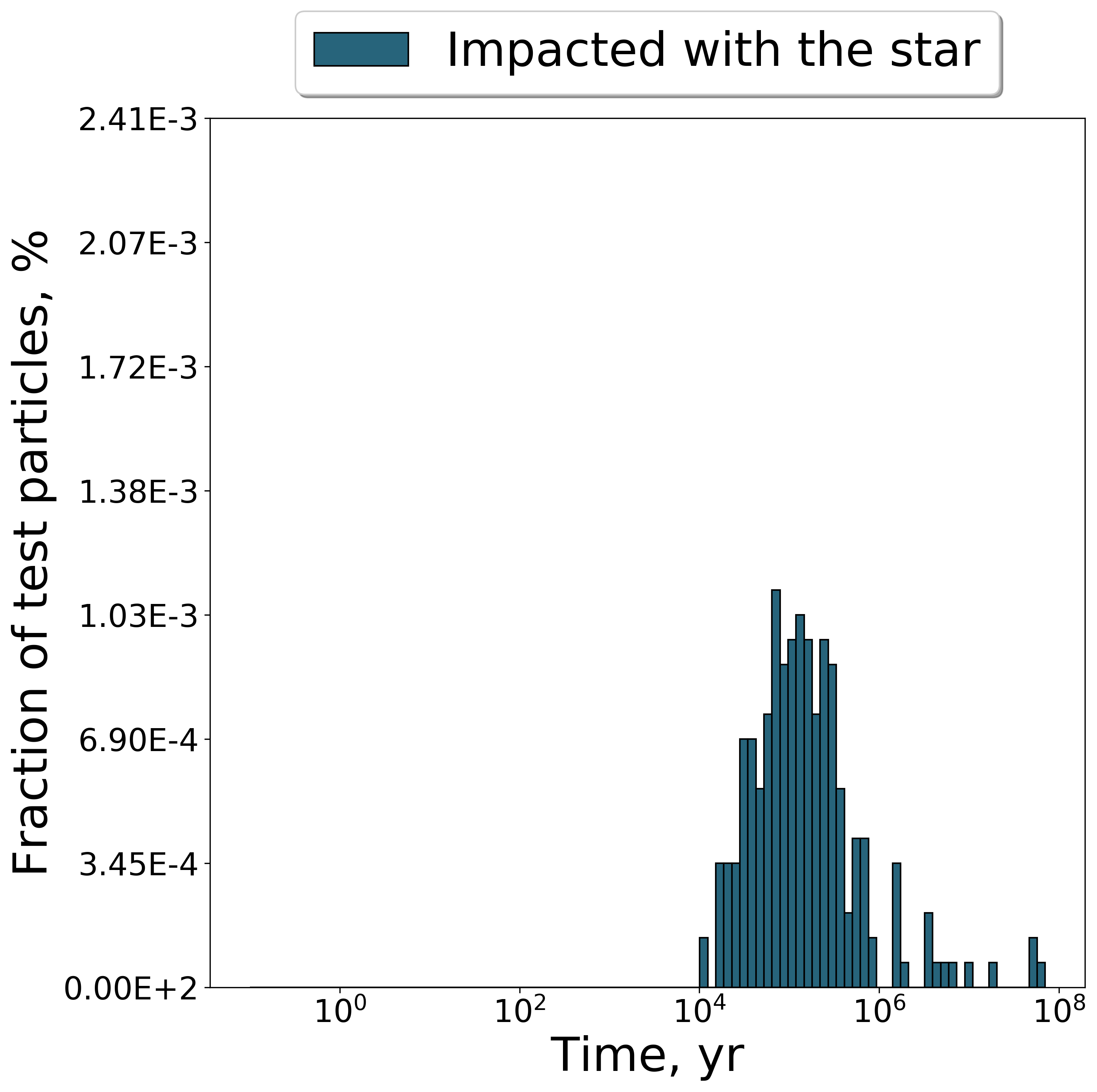}}
{\includegraphics[width=.49\linewidth]{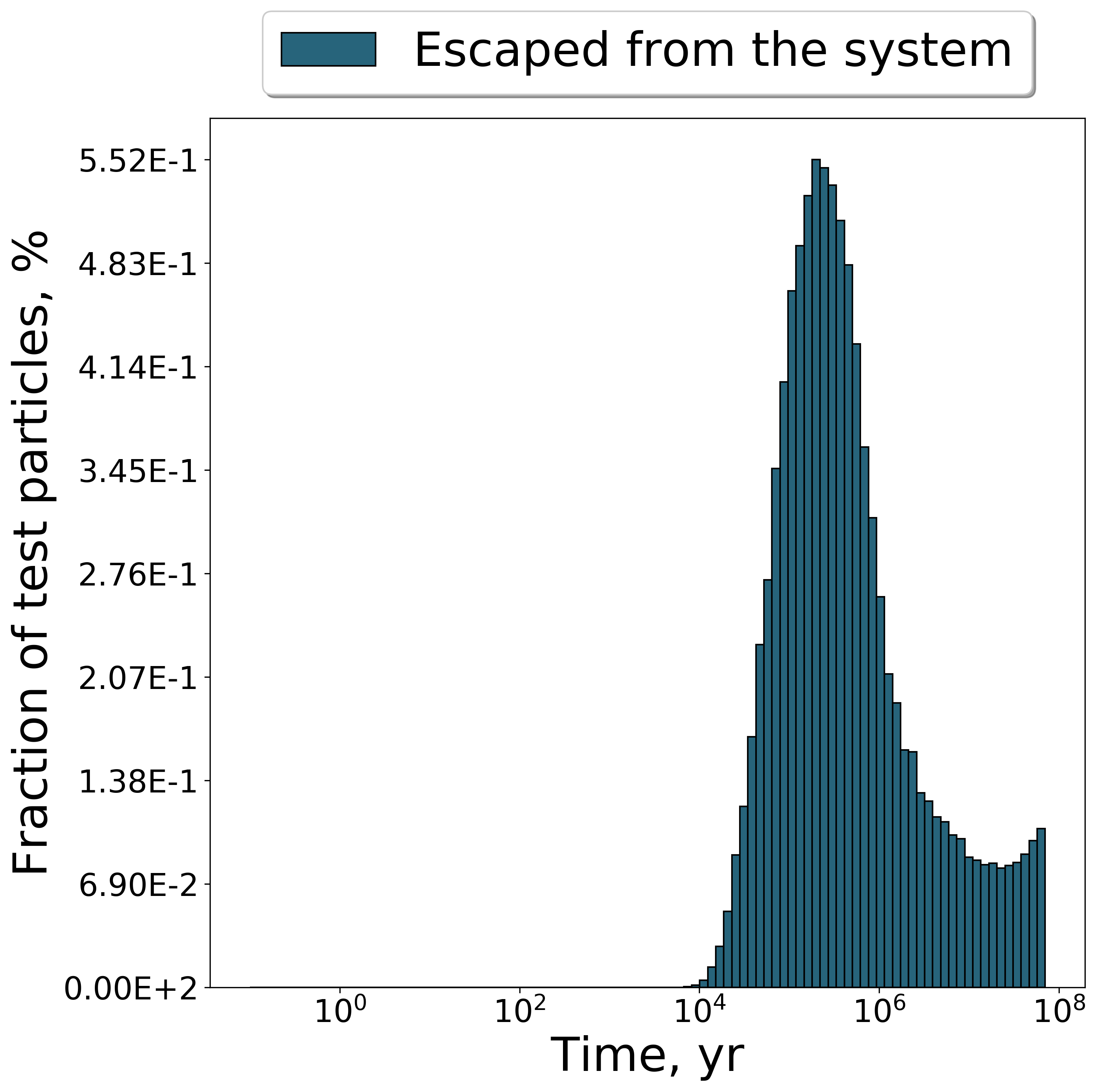}}
\caption{Fraction of planet impactors representing minor bodies originating from the outer belt after 10, 20, 30, ..., 70~Myr, respectively.}
\label{fig5.3.2:2}
\end{figure}

\begin{figure}
\centering
{\includegraphics[width=.49\linewidth]{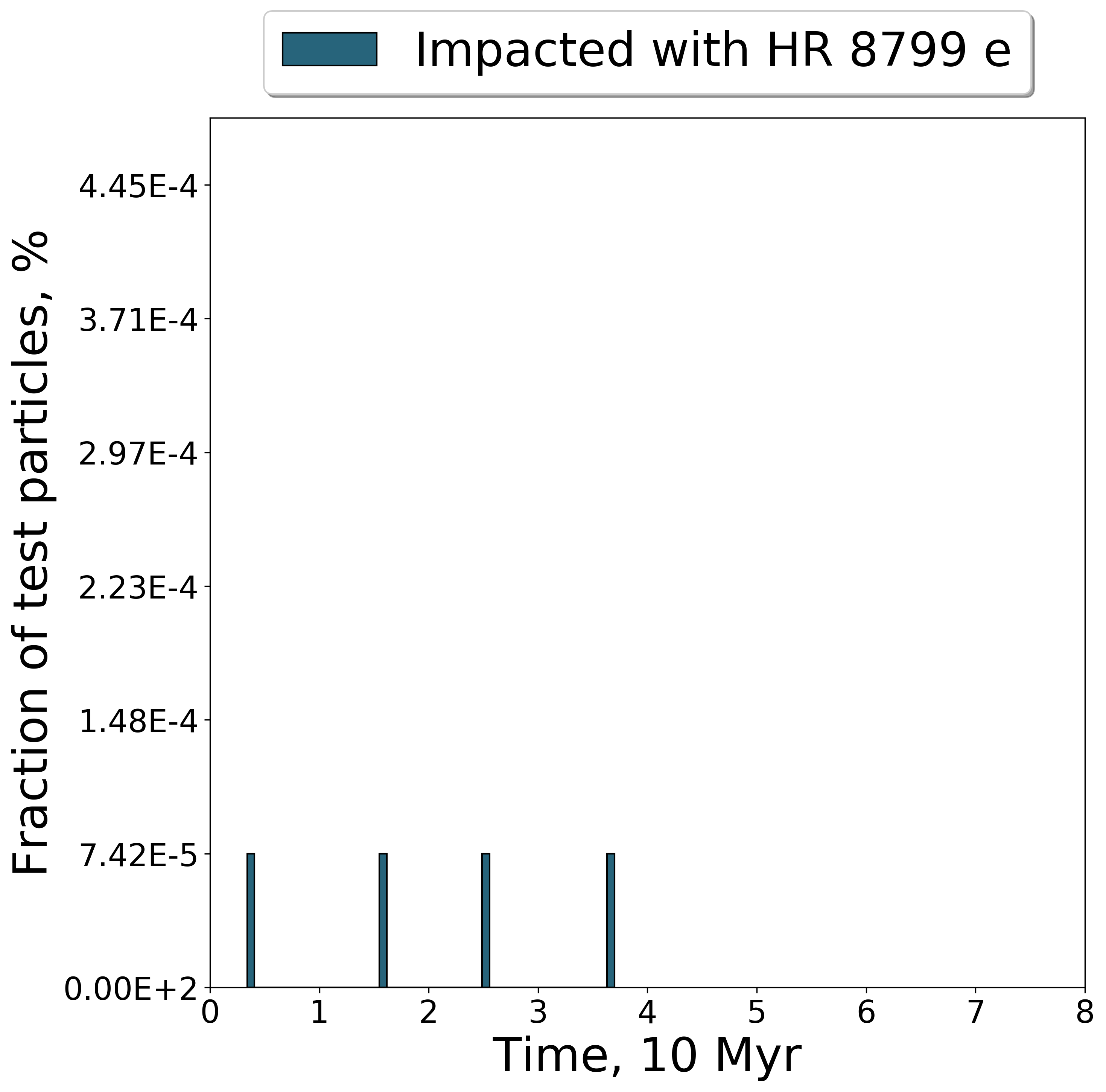}}
{\includegraphics[width=.49\linewidth]{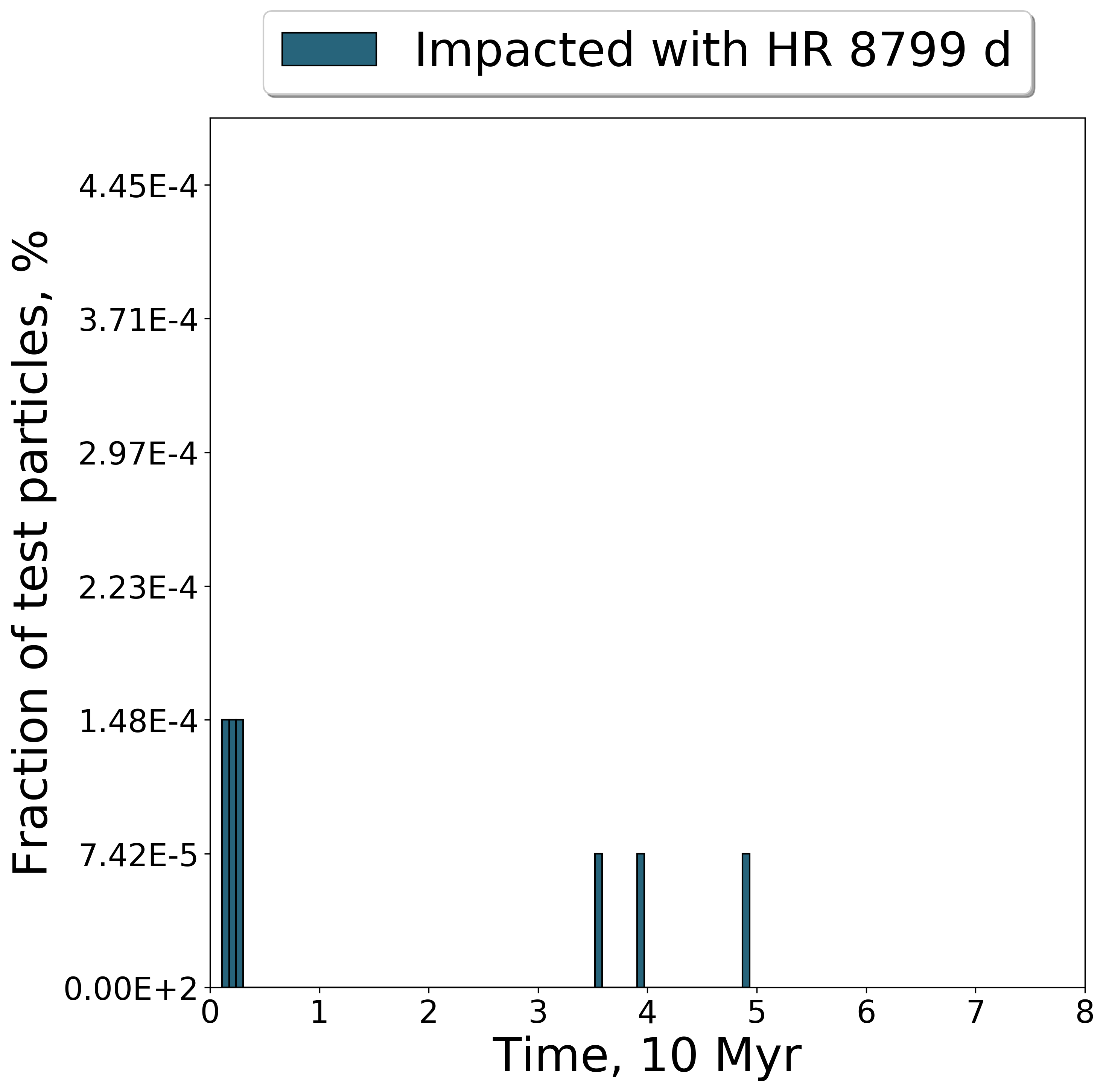}}
{\includegraphics[width=.49\linewidth]{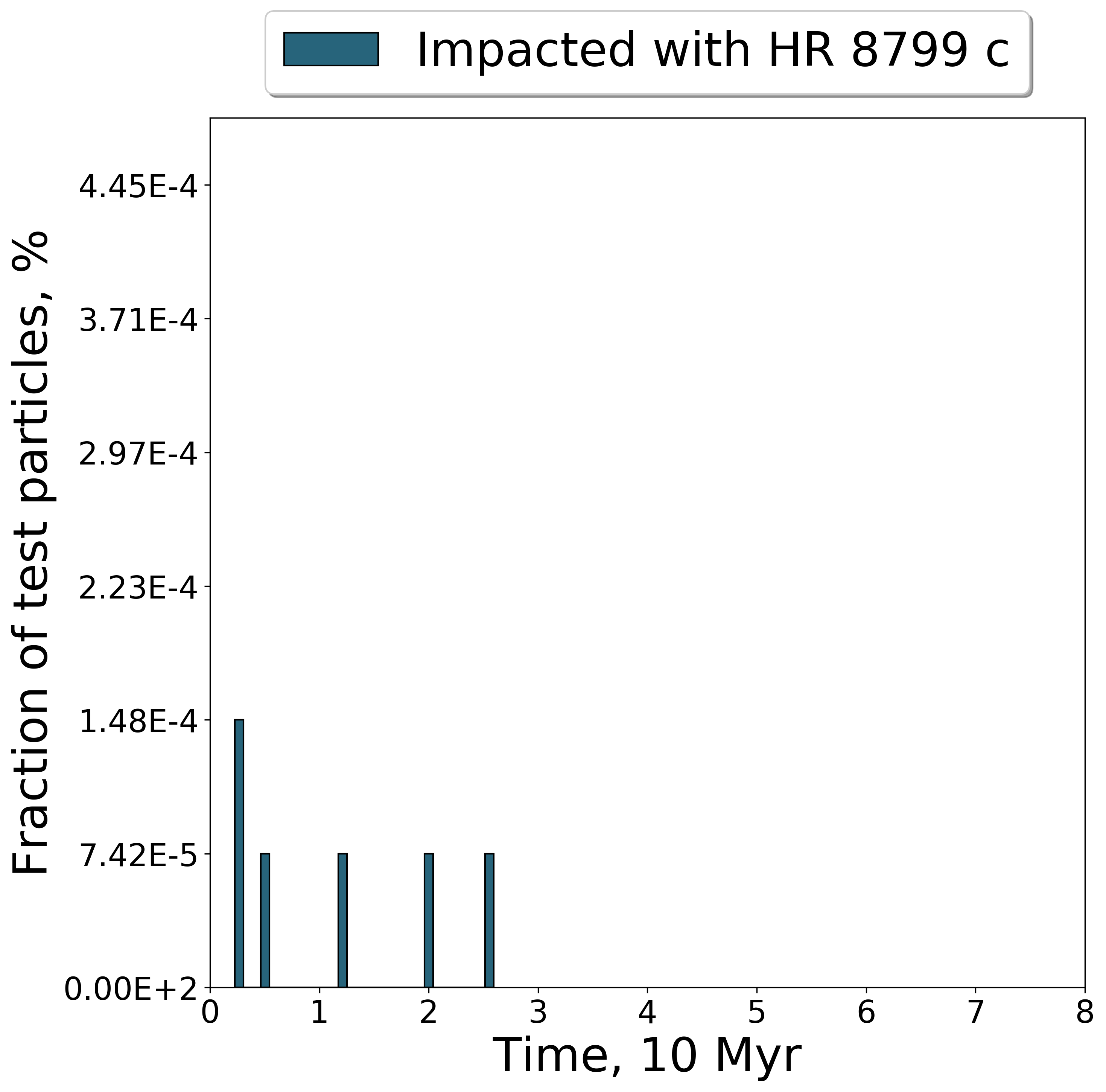}}
{\includegraphics[width=.49\linewidth]{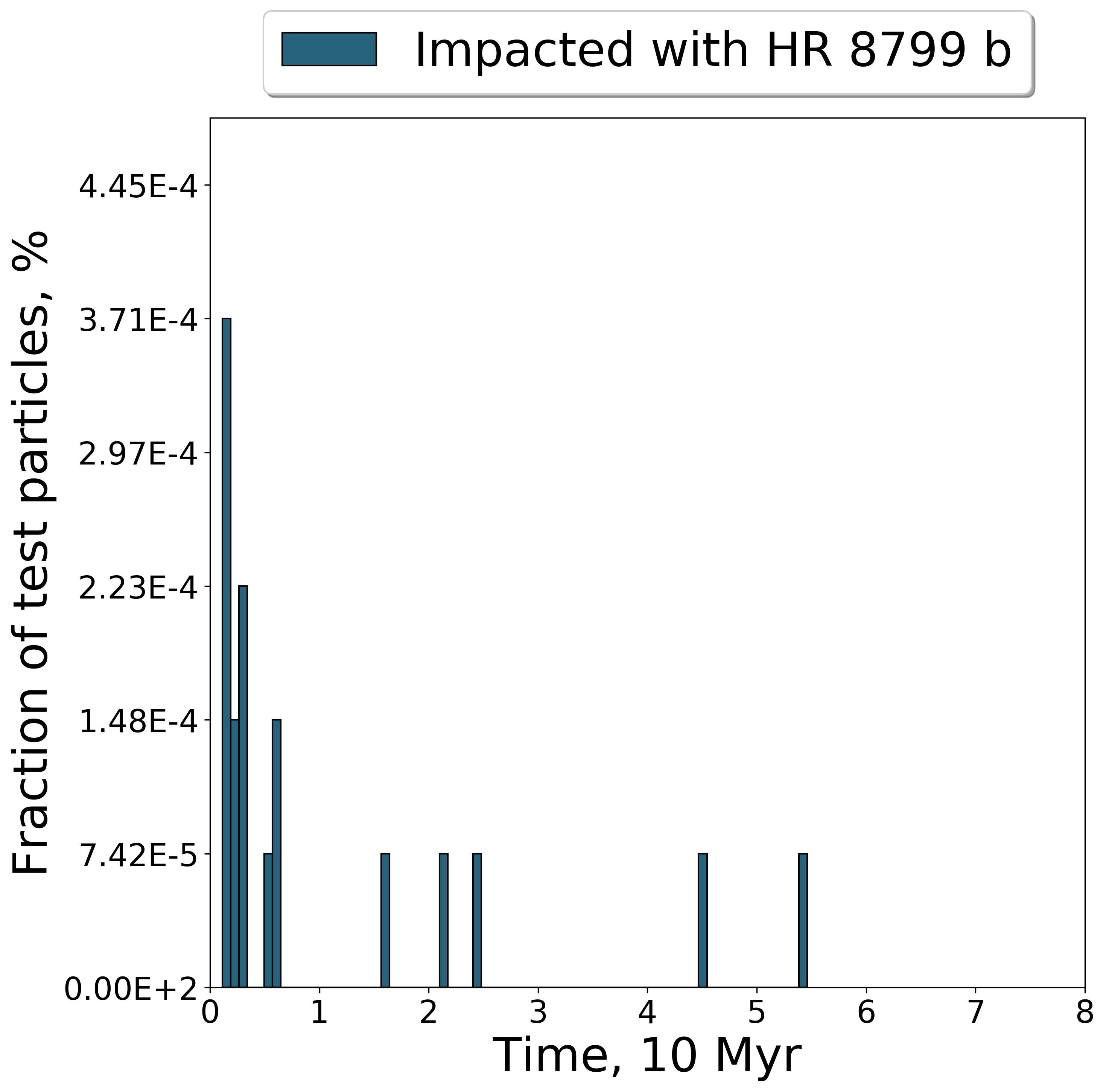}}
{\includegraphics[width=.49\linewidth]{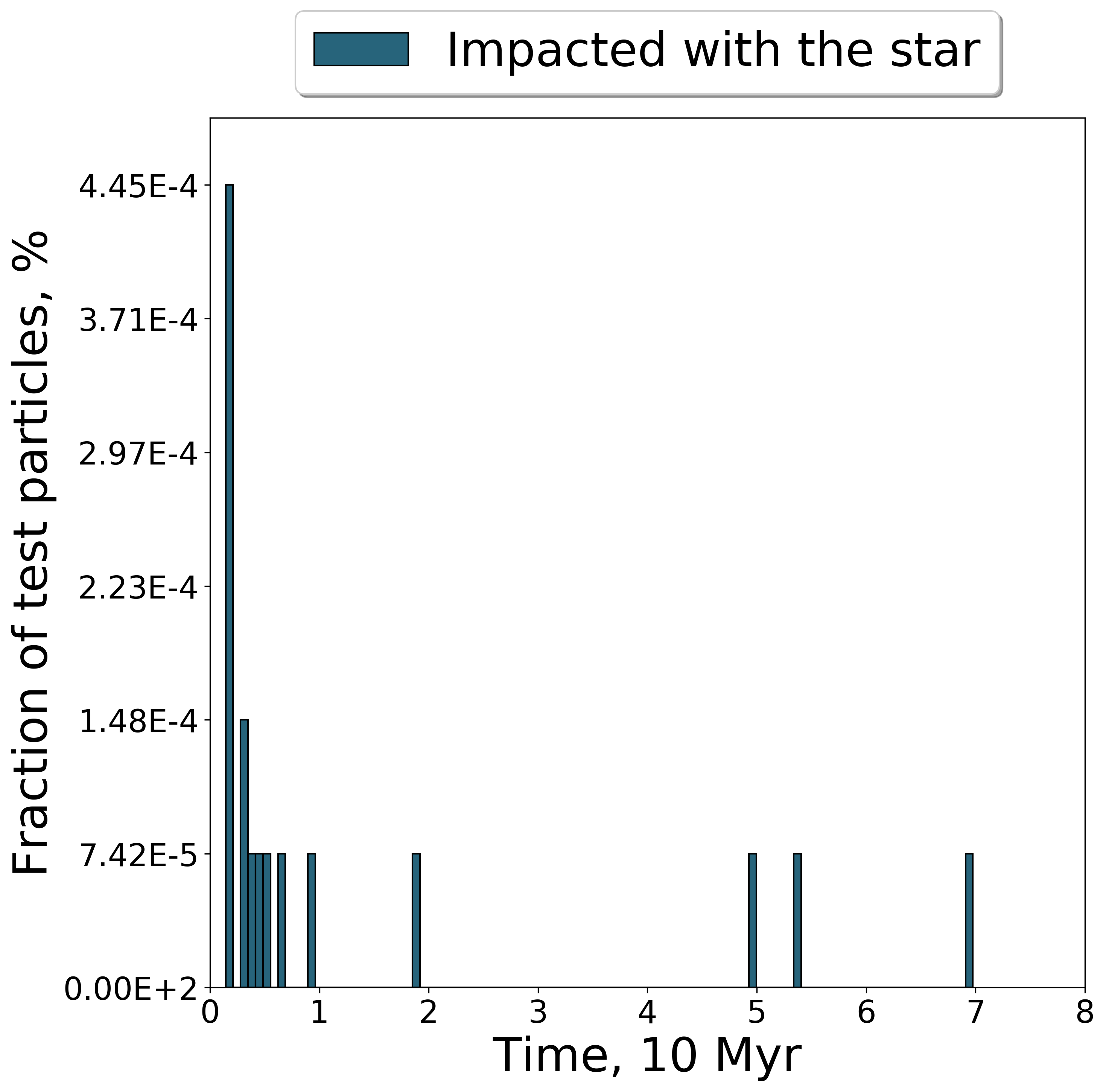}}
{\includegraphics[width=.49\linewidth]{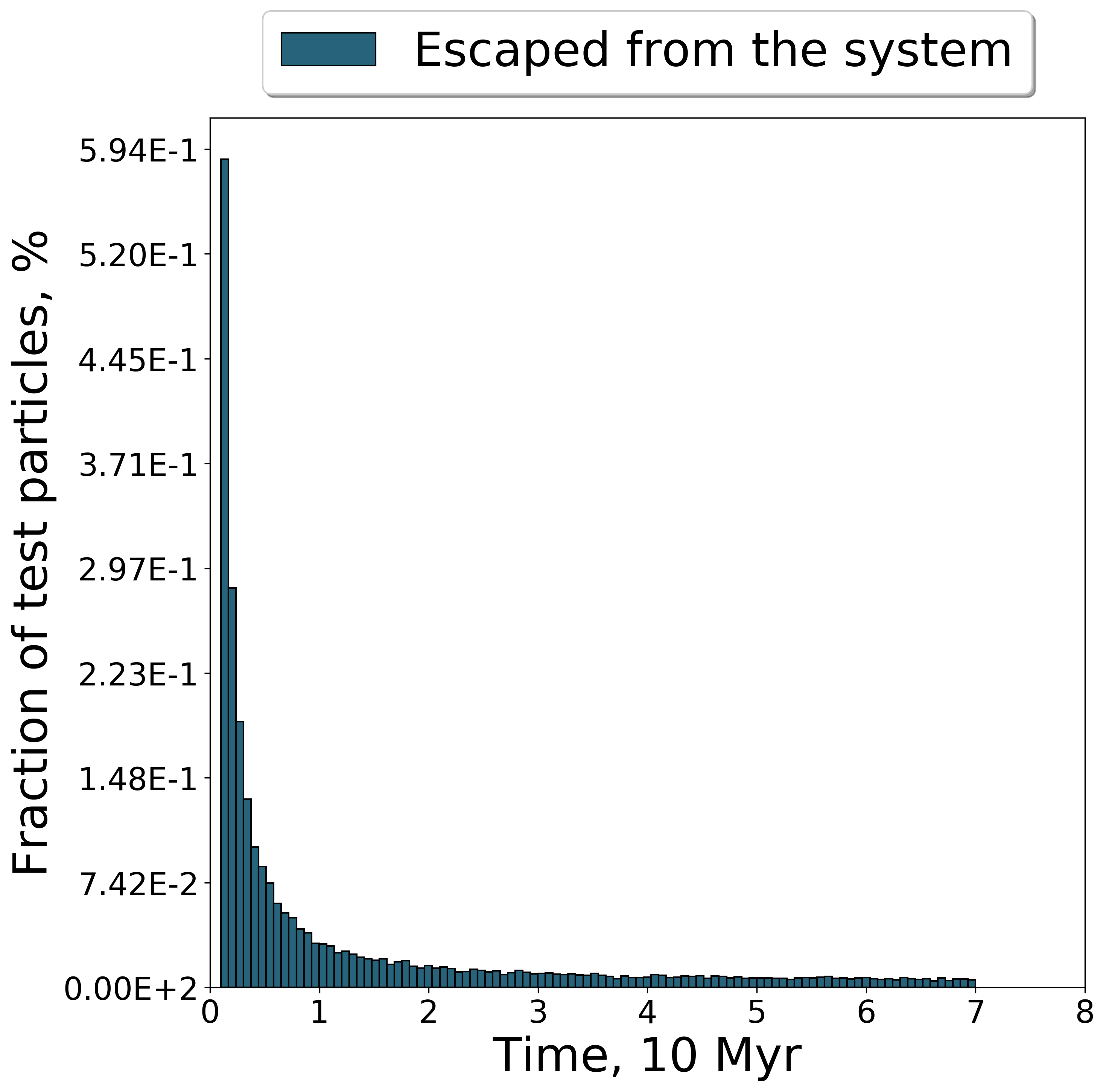}}
\caption{Fraction of planet impactors representing minor bodies which happened after the first 1~Myr of the simulations, after the system reached steady state.}
\label{fig5.3.2:3}
\end{figure}

In analogy with the inner belt, the structure of the outer belt is clearly visible by the time of the first snapshot at 6 Myr due to the interaction between the test particles and the planets as shown in Fig.~\ref{fig5.3.2:1}. After 70~Myr, 0.0143\% of test particles collided with the star and 9.4581\% of test particles were ejected from the system. As seen in Table~\ref{tbl5.3.1:1}, the outermost planet HR~8799~b suffered impacts from 0.0410\% of test particles, HR~8799~c from 0.0077\%, HR~8799~d from 0.0050\% and the innermost HR~8799~e from 0.0037\%. Fig.~\ref{fig5.3.2:2} shows that most impacts occurred in the first 1 Myr. The majority of the impactors is produced by the test particles initialised on highly unstable orbits. Steady state is reached around 1~Myr. We are interested in those impacts that occur after steady state was reached, see Fig.~\ref{fig5.3.2:3}. The structure of the outer belt in our simulations is the same as in previous work \citep[compare to Fig.~2 in][]{Read2018}. The range of orbits within the initial inner edge of the outer belt, 69.1~AU, and 100~AU is depleted by the end of the simulations. At the location of the 2:1 mean motion resonance with the outermost planet, a broad gap is formed. Also the simulations produce a few HR~8799~b "Trojans" orbiting at the same semi-major axis as the planet. Unlike \citeauthor{Read2018} we resolve and analyse impacts between planets and test particles, see \S\ref{sec5.4:VolatileDeliveryRates}.

%__________________________________________________________________

\subsubsection{Impact rates}\label{sec5.3.3:Impact rates}

The number of test particles representing minor bodies impacting the planets from both belts are shown in Table~\ref{tbl5.3.1:1}. We only consider those impacts that occurred after the system reached steady state, 1-70~Myr. The innermost planet receives 7.2 and 0.06 impacts/Myr from the inner and the outer belts. Planet d receives 0.6 and 0.1 impacts/Myr, while planet c 0.3 and 0.09 impacts/Myr. The outermost planet receives 0.07 and 0.3 impacts/Myr.

The total number of test particles, $N_\textrm{tp}$, in the inner belt is 544,659 and 1,347,075 in the outer belt. We note that these values represent the total number of test particles in either belt after the first 1~Myr when the system reached steady state and differ from the 650,000 test particles and 1,450,000 per inner and outer belts initialised at the start of the simulations.

%-------------------------------------------------------------------

\subsection{Dust}\label{subsec:SimulationResultsDust}
In our numerical simulations of two distinctive dust populations originating in the inner and outer belts, we recorded the dust dynamical evolution in the HR~8799 system. Since our simulations do not include effects of dust-dust collisions that are estimated in the post-processing, we obtained the complete dynamical history of each simulated particle. These simulations resulted in 150 million unique records of orbital elements for the outer belt dust population, and 188 million records for the inner belt. Due to their significantly different source regions and size range, we describe each population separately in the following paragraphs.

The dynamical pathways of the outer belt dust ($10~\mu\mathrm{m} <D<1000~\mu$m) are mostly driven by a slow migration into outer mean-motion resonances with HR~8799~b, influence of secular resonances and planetary close-encounters that lead to scattering of dust into outer space ($R>10,000$~AU). Only less than 5\% of dust is able to migrate below $R<3$~AU assuming that there are no dust-dust collisions. The average collision probability $P$ of particles migrating via PR drag from the outer belt is $P\approx 10^{-10}$ per year for HR~8799~b, and decreases by a factor of 2 for each subsequent planet. We calculate $P$ adopting the collision probability between orbiting objects from \citet{Kessler_1981}. In reality, the mass of the outer dust ring is 18 million times higher than the mass of the Zodiacal Cloud \citep{Nesvorny2011a} $4\times10^{19}$ g vs. $0.12\,\textrm{M}_\Earth = 7.2 \times 10^{26}$ g, and therefore destructive dust-dust collisions happen on orders of magnitude shorter timescales than the PR drag dynamical timescales, as suggested in \citet{Wyatt_2005} for young debris discs like HR~8799. 

\begin{figure}
\centering
{\includegraphics[width=.99\linewidth]{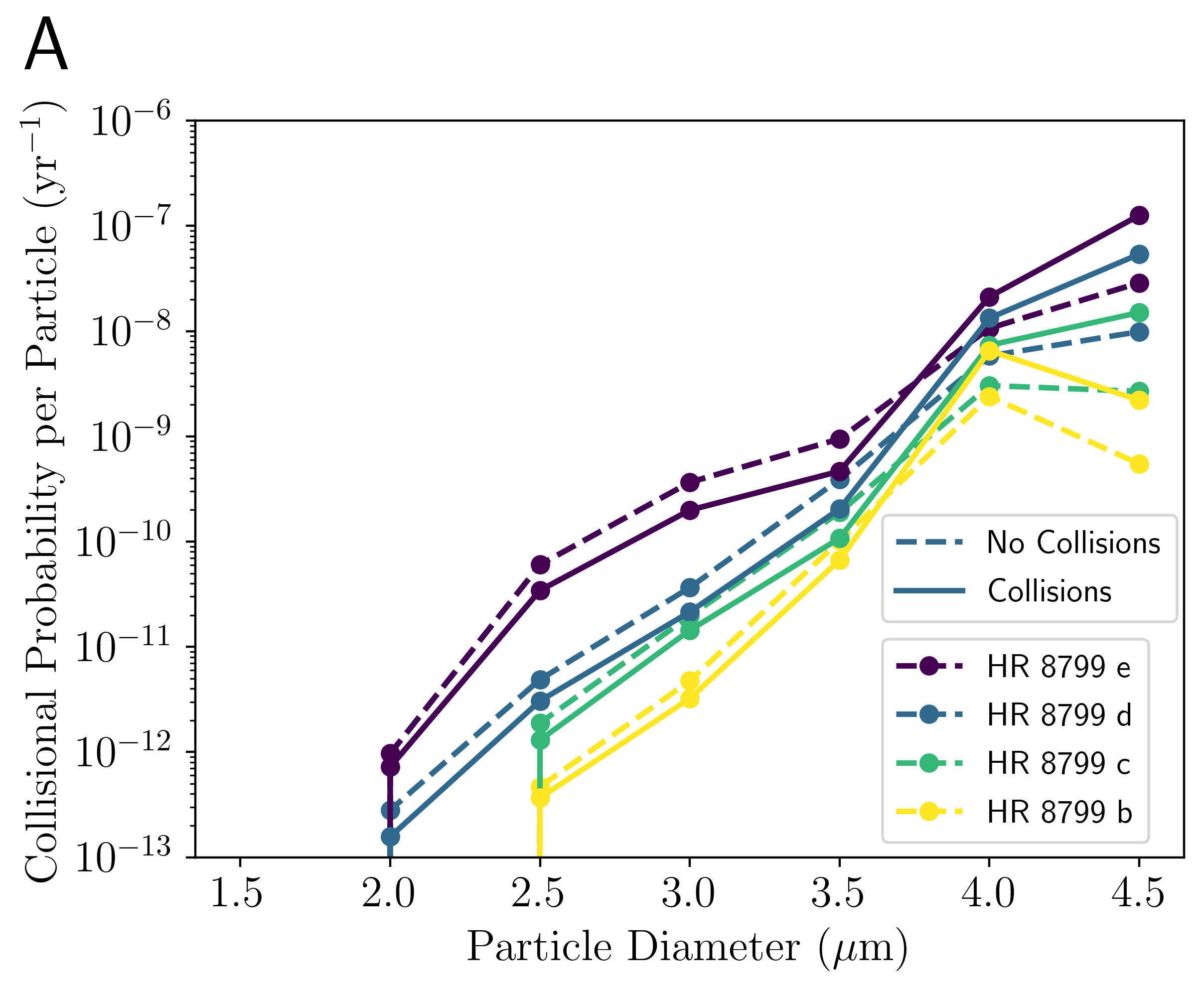}}
\vspace{\fill}
{\includegraphics[width=.99\linewidth]{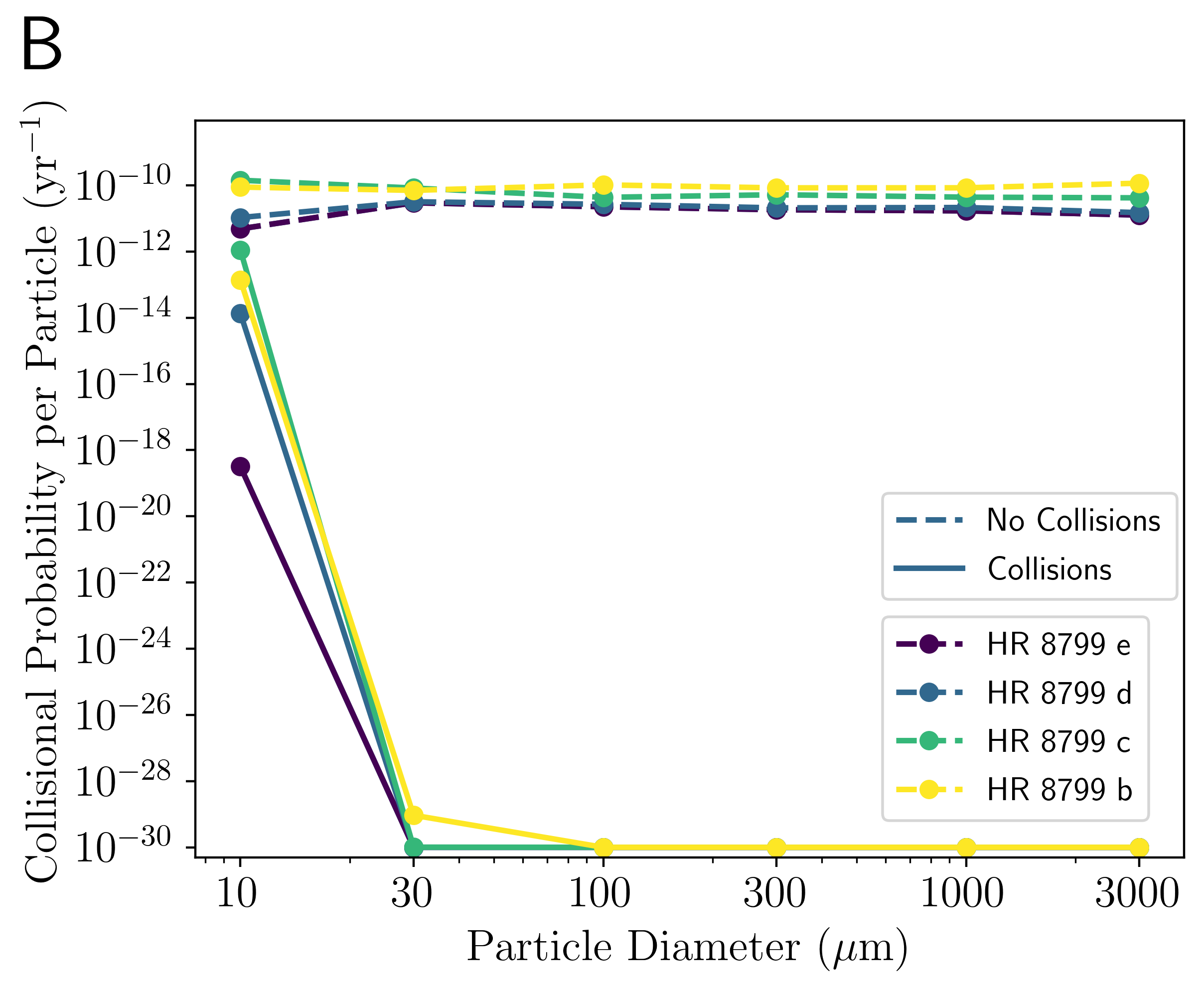}}
\caption{Panel A: Average collisional probability with each HR~8799 planet for 7 different particle diameters originating in the inner belt. Each colour corresponds to a different planet, where the dashed lines are assuming no particle-particle collisions, while solid lines are including collisions (see the main text for details). The units are particle impacts onto the entire planet surface per year. Panel B: The same as Panel A but now for the outer belt particles and 6 different particle diameters. In order to show the difference between scenarios with and without particle-particle collisions, we added a constant value of $P = 10^{-30}$ to all data points in Panel B. Except for $D=30~\mu$m impacts onto planet HR~8799~b all particles with $D\ge30~\mu$m have $P \ll 10^{-30}$ when particle-particle collisions are included.}
\label{fig5.3.2:4}
\end{figure}

To check the effects of dust-dust collisions in the HR~8799 system, let us assume that the HR~8799 outer belt dust has a similar shape as the Zodiacal Cloud, and use the collisional balance estimates from \citet{Grun1985}. We multiply the density of the cloud by a factor of 18 million and scale it to a distance of 100~AU. This results in PR drag driven migration, where only the smallest $D=10\mu$m particles can reach HR~8799~b ($a=69.1$~AU) with an impact probability $P=10^{-13}$ per year. For any dust larger than $D=10~\mu$m the impact probability even with the outermost planets drops quickly to zero for $D<21.5\,\mu$m; $P<10^{-30}$ per year as shown in the Figure's~\ref{fig5.3.2:4} panel~B. From this we conclude, that shortly after the outer belt is shaped by the gravitational interaction with planets (Fig.~\ref{fig5.3.1:2} - 6 Myr), the dust contribution from the outer belt is negligible due to the overwhelming dust-dust collisional rate. This is also independently confirmed in \citet{Wyatt_2005}, who concludes that for all resolved debris discs with ages $<100$ Myr the effects of PR drag are insignificant. The source populations of dust are far behind the outermost planets and thus particles cannot reach planet crossing orbits before their collisional demise. The dust fragments are either blown out of the system or subject to more collisions. The PR timescale for a $D=10\,\mu$m particle starting at $a=100$~AU on a circular orbit to reach planet HR~8799~b is 7.2~Myr \citep{Wyatt_2005}, for our assumed bulk density of $\rho=2$ g cm$^{-3}$. We can thus effectively disregard any possible delivery of material from the outer dust belt on any of the four planets.

The situation is vastly different for the inner dust population due to the particle size range, $1.5~\mu\mathrm{m} < D < 4.5~\mu$m. These sizes are very close to the blow out regime, $\beta>0.5$ for circular orbits of their parent bodies, where $\beta \approx (1.15\times10^{-3}/\rho D) (L_\star/\textrm{L}_\odot)(\textrm{M}_\odot/M_\star)$ \citep{Wyatt_etal_1999}, where $\rho$ and $D$ are in SI units. The critical diameter when $\beta=0.5$ is $D_\star=3.85~\mu$m for $\rho=2$ kg m$^{-3}$. Since our parent bodies have non-zero eccentricity, particles smaller than $D_\star=3.85~\mu$m that are ejected close to aphelion can retain bound orbits. Particles with $D<3.0~\mu$m have almost zero impact probability with all four planets ($P<10^{-13}$ per year as shown in the Figure's~\ref{fig5.3.2:4} panel~A.), thus we conclude they do not deliver any significant portion of material to any planet. Dust particles with $3.5<D<4.5\mu$m have the highest impact probability ($P\sim 10^{-8}$ per year) with the innermost planet, and the impact probability decreases by a factor of 2 for each subsequent planet further from HR~8799. The total mass of dust in the inner belt is $M_{\textrm{dustInner}}=1.1\times10^{-6}\,\textrm{M}_\Earth$, so at the impact rate we calculated here, the HR~8799 planets would in 100~Myr accrete $<10^{-6}\,\textrm{M}_\Earth$, which is $~9$ orders of magnitude less than their estimated mass.

To summarise, our numerical integrations show that both the inner and outer dust populations do not significantly contribute to material delivery onto any of the four known planets orbiting HR~8799. This is unlikely to change until the rate of collisions/optical depth of the outer belt drops significantly and allows inward migration of dust via PR drag.

Additionally, it is important to highlight that the dust size ranges, $1.5-4.5~\mu$m for the inner belt and $10-1000~\mu$m for the outer belt, have been chosen based on the existing observations of the system. Such discontinuous size ranges between these two populations can be due to the observational bias because Spitzer is sensitive to small grains and ALMA to large ones. Since we do not have observational constraints on the distribution of the larger grains in the inner belt and the smaller grains in the outer belt, we do not try to add these size ranges in our simulations. 

%__________________________________________________________________

\section{Delivery rates}\label{sec5.4:VolatileDeliveryRates}

In Section~\ref{sec5.3:SimulationResults} we have derived impact rates.  Here, we use these to estimate the corresponding delivery rates of volatile (water and organics) and refractory (metals and silicates) material to the four planets. In Section~\ref{subsec:SimulationResultsDust} we have demonstrated that the dust populations of both belts do not significantly contribute to the delivery of material to the giant planets, therefore the delivery rates calculated in this section refer only to the delivery by minor bodies.

The volatile delivery rate $R_\textrm{vol}$ and the refractory delivery rate $R_\textrm{refr}$, for each planet, can be expressed as \citep[see][Equation 3]{Schwarz2018}:

\begin{equation}
     R_\textrm{vol} = \frac{M_\textrm{belt} \times f_\textrm{vol} \times N_\textrm{imp}}{N_\textrm{tp}\,\tau_\textrm{sim}},
\label{eq5.4:1}
\end{equation}

\begin{equation}
     R_\textrm{refr} = \frac{M_\textrm{belt} \times f_\textrm{refr} \times N_\textrm{imp}}{N_\textrm{tp}\,\tau_\textrm{sim}},
\label{eq5.4:2}
\end{equation}

where $M_\textrm{belt}$ is the total mass of the belt, $f_\textrm{vol}$ and $f_\textrm{refr}$ are the mass fraction of volatiles and refractories therein, $N_\textrm{imp}$ is the number of impacts on a planet from the corresponding belt, $N_\textrm{tp}$ is the number of test particles in the corresponding belt after the first 1~Myr of the simulation and $\tau_\textrm{sim}$ is the simulation time. The simulation time $\tau_\textrm{sim}$ is 69~Myr and not 70~Myr since we do not consider the impacts that happened in the first 1~Myr of the simulations. To estimate values of $M_\textrm{belt}$, $f_\textrm{vol}$ and $f_\textrm{refr}$ for the inner belt we use the Main Asteroid belt as a proxy and the Kuiper belt for the outer belt.

The values of $N_\textrm{imp}$, $N_\textrm{tp}$, $\tau_\textrm{sim}$ have been determined in \S\ref{sec5.3.3:Impact rates}. In this section, we estimate $M_\textrm{belt}$, $f_\textrm{vol}$ and $f_\textrm{refr}$.

%-------------------------------------------------------------------

\subsection{Belt masses}

The total dust mass of either belt was determined by dust modelling based on infrared observations. According to \citet{Su2009} the total mass of the $1.5\,-\,4.5\,\mu$m sized dust grains in the inner belt is $M_\textrm{dustInner} = 1.1 \, \times 10^{-6} \textrm{M}_\Earth$ and the mass of the $10\,-1000\,\mu$m sized dust grains in the outer belt is $M_\textrm{dustOuter} = 0.12 \, \textrm{M}_\Earth$. \citet{Geiler2019} created a disc model that would explain Herschel and ALMA observations of the outer belt. Their "preferred" model predicts $M_\textrm{plOuter} = 134 \, \textrm{M}_\Earth$ as the total mass of the outer belt, assuming a size-frequency distribution of minor bodies up to a maximum diameter of 100~km. We adopt this value for our further calculations. 

For the total minor body mass of the inner belt, however, there are no published estimates that we are aware of. We estimate the total belt mass (minor bodies only) based on the dust mass determined by \citet{Su2009}. In doing so, we assume a power law describing the differential size-frequency distribution $n(a)$ (diameter $a$) of a colliding steady state population of bodies:

\begin{equation}
    n(a) \, \propto \, a^{-q},
\label{eq5.4:3}
\end{equation}

where the power-law index $q\,=\,3.5$ \citep{Dohnanyi1969,Bottke2015}. The total mass of the inner belt for minor bodies can be calculated by integrating over the size distribution from the minimum to maximum dust sizes (from $a_\textrm{minD} = 1.5$ to $a_\textrm{maxD} = 4.5\,\mu$m as observed by \citet{Su2009}) and minor body sizes ($a_\textrm{minD} = 1$~m to $a_\textrm{maxD} = 1000$~km, based on the size distribution of the Main Asteroid Belt):

\begin{equation}
    m(<a_\textrm{max}) \, = \, \frac{\pi}{6} \, \int_{a_\textrm{min}}^{a_\textrm{max}} \, n(a) \, \rho \, a^3 \, da \, \propto \, \sqrt{a} \, |_{a_\textrm{min}}^{a_\textrm{max}}. 
\label{eq5.4:4}
\end{equation}

Normalising the total inner belt mass for minor bodies by the total inner belt mass for dust particles gives:

\begin{equation}
    \frac{M_\textrm{plInner}}{M_\textrm{dustInner}} = \frac{\sqrt{a_\textrm{maxA}} - \sqrt{a_\textrm{minA}}}{\sqrt{a_\textrm{maxD}} - \sqrt{a_\textrm{minD}}}. 
\label{eq5.4:5}
\end{equation}

In doing so, we found the total mass of the inner belt to be $M_\textrm{plInner} = 1.23 \, \textrm{M}_\Earth$. For comparison, the estimated dust mass of the inner Solar System is $7 \times 10^{-6} \, \textrm{M}_\Earth$ \citep{Nesvorny2011a} and the estimated mass of the Main Asteroid Belt is $0.0004 \, \textrm{M}_\Earth$ \citep{DePater2015}. The planetesimal to dust mass ratio for the inner belt of the HR~8799 is $10^6$ compared to $10^3$ for the Solar System.

%-------------------------------------------------------------------

\subsection{Volatile content}

While it is reasonable to assume that the debris belts in HR~8799 contain some volatile material, no observational constraints are available. We estimate the volatile content $f_\textrm{vol}$ of the inner and outer debris belt, basing ourselves on knowledge from the Solar System. 

Solar System asteroids are classified in several taxonomic types. Only members of one such type, the so-called C-type asteroids, show appreciable amounts of volatiles. About $33$\% of all asteroids belong to this type \citep{DeMeo2013}. The water content of the C-type asteroids is $\approx10\%$ by mass and their carbon content is $\approx2\%$ by mass \citep{EOE2007,Sephton2002,Sephton2014}. Therefore, we adopt $0.33 \times (0.10 + 0.02) = 0.04$ as the $f_\textrm{vol}$ value for the inner belt of the HR~8799 system.

In contrast, the Kuiper Belt objects all are formed beyond the snow line, which means that on average all those objects are expected to be volatile rich. However, the ice fraction measured in KBOs spans the full range from 0 to 1 \citep{Brown2012}. Since the range of the volatile fraction is rather wide we adopted an average value of 0.5 as the $f_\textrm{vol}$ for the HR~8799 outer belt. The same value was adopted by \citet{Ciesla2015} and \citet{Schwarz2018} in their studies of volatile delivery. 

After inserting all necessary values into Eq.~\ref{eq5.4:1} we calculate the volatile delivery rates, as shown in Table~\ref{tbl5.4:1} from the minor bodies of the inner and the outer belts to the four giant planets. The outer belt delivers an order of magnitude more volatiles to the planets despite the smaller total number of impacts. Especially for the outer belt our simulations yield few impacts resulting in large Poisson noise, up to 50\%, leading to large uncertainties in delivery rates. 

\begin{table*}
\caption{Volatile and refractory delivery rates in Earth masses per Myr, $\textrm{M}_\Earth$/Myr, from the minor bodies of the inner belt and the outer belt to the planets HR~8799~e, HR~8799~d, HR~8799~c, and HR~8799~b.}
\label{tbl5.4:1}
\centering
\begin{tabular}{rrrrr}
\hline
belt & e & d & c & b \\
\hline
\multicolumn{5}{c}{volatiles} \\
inner & ${6.4\pm0.3\,\times\,10^{-7}}$ & ${4.9\pm0.8\,\times\,10^{-8}}$ &  ${2.7\pm0.6\,\times\,10^{-8}}$ &  ${6.5\pm2.9\,\times\,10^{-9}}$ \\
outer & ${2.9\pm1.4\,\times\,10^{-6}}$ & ${6.5\pm2.2\,\times\,10^{-6}}$ &  ${4.3\pm1.8\,\times\,10^{-6}}$ &  ${1.3\pm0.3\,\times\,10^{-5}}$ \\
\multicolumn{5}{c}{refractories} \\
inner & ${1.6\pm0.1\,\times\,10^{-5}}$ &  ${1.2\pm0.2\,\times\,10^{-6}}$ &  ${6.6\pm1.4\,\times\,10^{-7}}$ &  ${1.6\pm0.7\,\times\,10^{-7}}$ \\ 
outer & ${2.9\pm1.4\,\times\,10^{-6}}$ & ${6.5\pm2.2\,\times\,10^{-6}}$ &  ${4.3\pm1.8\,\times\,10^{-6}}$ &  ${1.3\pm0.3\,\times\,10^{-5}}$ \\
\hline
\end{tabular}
\end{table*}

%-------------------------------------------------------------------

\subsection{Refractory content}

Like for volatiles, we base our estimates of refractory-material content on our knowledge of the Solar System. In particular, we assume that any material not counted as volatile is refractory; in practice, the refractory component is likely to be dominated by silicates and metals.  We therefore adopt a refractory content $f_\textrm{refr}$ of 0.96 for the inner belt and 0.5 for the outer belt, respectively. The resulting delivery rates are presented in Table~\ref{tbl5.4:1}.

%__________________________________________________________________

\section{Discussion}\label{sec5.5:Discussions}

As Table \ref{tbl5.4:1} shows, volatiles and refractories are delivered from both belts to all four planets by minor bodies. The inner belt delivers to the planets ${7.3 \times 10^{-7}\,\textrm{M}_\Earth}$ of volatile material per Myr and ${1.7 \times 10^{-5}\,\textrm{M}_\Earth}$ of refractory material per Myr. The outer belt delivers ${2.7 \times 10^{-5}\,\textrm{M}_\Earth}$ of volatiles per Myr and ${2.7 \times 10^{-5}\,\textrm{M}_\Earth}$ of refractories per Myr.

In terms of the impact rates the inner belt dominates the outer belt except for planet HR~8799~b. For example, planet HR~8799~d experiences 0.6~impacts/Myr from the inner belt and 0.1~impacts/Myr from the outer belt. However, the outer belt is much more massive than the inner belt, two orders of magnitude, and its volatile fraction is an order of magnitude higher than for the inner belt. As a result, the volatile delivery to the planets is dominated by the outer belt. On the other hand, the refractory fraction of the inner belt is almost twice as large as for the outer belt, which leads to similar refractory delivery rates from both belts.

Over the course of 69~Myr the four giant planets receive between ${2\pm1\times10^{-4}\,\textrm{M}_\Earth}$ and ${9\pm2\times10^{-4}\,\textrm{M}_\Earth}$ of volatile material and between ${3\pm1\times10^{-4}\,\textrm{M}_\Earth}$ and ${13\pm2\times10^{-4}\,\textrm{M}_\Earth}$ of refractory material. The uncertainty is dominated by the Poisson noise in the simulated number of impacts, and by the planetesimal masses of the belts. These total volatile and refractory fluxes are small compared to the total planet mass, which is $10^6 - 10^7$ times larger. Such an amount of volatile and/or refractory infall may be detectable in the upper layers of the planetary atmospheres. For example, gases with mixing ratios of $\sim 10^{-6}$ and even lower are often detectable, depending on species \citep{DePater2015}.

Since the four giants HR~8799~e, d, c, b are beyond the snow line (and presumably formed there), we expect them to be born volatile-rich. Any future detection of volatiles would therefore not necessarily imply delivery through impacts. Silicates or other refractory material would be more diagnostic in this regard. Past observations of Jupiter after the impact of comet Shoemaker-Levy 9 and smaller objects may be used as a proxy for post impact enrichment observations of the giant planets in the HR~8799 system \citep{ATREYA1999,Fletcher2010}. 

The HR~8799 system may contain terrestrial planets, which presumably formed dry within the snow line. Volatile delivery from the belts may be of astrobiological importance for those. The same is true, probably, for possible terrestrial planets around other stars with minor body belt analogues. Given the lower mass of terrestrial planets (an Earth analogue would be about 2,800 less massive than HR~8799~e), the relative contribution of impactor material to the overall composition of the planet and its atmosphere could be much higher. Volatiles derived by impacts could be enough to explain an Earth-like atmosphere mass: Earth's atmosphere contributes about $10^{-6}$ of Earth's total mass.  We caution, however, that we did not (yet) model the impactor flux on terrestrial planets in the HR~8799 system.

We also analysed a potential transport of material from the dust belts to the four giant planets. For the outer belt model only meteoroids with $D=10~\mu$m deliver material to HR~8799 planets with a non-zero probability due to their short PR timescales. We can extrapolate this behaviour to particles smaller than that, which are migrating even faster due to their larger $\beta$ values, and thus are less susceptible to destruction in mutual meteoroid collisions. However, too small particles, $D<3~\mu$m, are immediately blown out of the system, thus only a narrow size range $D = 3-10~\mu$m remains. This fact greatly diminishes any possibility of any considerable mass transport from the outer belt to planets via micron sized dust. Particles with larger diameters ($D>1000~\mu$m) migrate inwards slower than $D<1000~\mu$m, are destroyed in dust-dust collisions, and should not deliver material to HR~8799 planets.

Inner belt particles with $D>4.5~\mu$m are released on stable orbits. These orbits are migrating inwards due to PR drag. Due to the effects of the radiation pressure the initial aphelion distance of released particles can extend beyond the orbits of the HR~8799 planets, however particles with $D>10.2~\mu$m, released from $a=8$~AU and $e=0.1$ bodies have aphelion distances $Q<15.4$~AU, meaning that orbits not crossing any of HR~8799 planets. It is possible to increase the semimajor axes of the inner belt dust via close encounters or secular resonances, however the dust-dust collisions are expected to act on much shorter timescales, and we do not believe that this outward particle migration is efficient enough to effectively deliver a significant portion of material onto the  HR~8799 planets. Moreover, the current constraints give a very low mass of the inner belt dust ($M_{\textrm{dustInner}}=1.1\times10^{-6} \textrm{M}_\Earth$), which further decreases the possibility of material transport from the inner belt via dust particles.

Our results show that the delivery process in the HR~8799 is dominated by minor bodies, while in the current Solar System  delivery processes are dominated by dust populations. This follows from the fact that HR~8799 is a young non depleted system thus the minor body influx on planets dominates, while Solar System is 4.5 billion years old, and most of the minor bodies are in stable reservoirs that do not intersect any planet.

Collisions between minor bodies in the outer and inner belts of the HR~8799 system produce dust, analogous to the zodiacal dust in the Solar system, that may be observable with the upcoming HOSTS (Hunt for Observable Signatures of Terrestrial Systems) survey \citep{Ertel2018a,Ertel2018b}. Possible observation of the exozodiacal dust around HR~8799 may provide insights on the existence of terrestrial planets in the system and on the nature of the belts.

%__________________________________________________________________

\section{Conclusions}\label{sec5.6:Conclusions}

We performed numerical simulations of the inner and outer debris belts in the exoplanetary system HR~8799 to calculate for the first time delivery rates from the belts to the four planets. Delivery rates were separated into a volatile and a refractory component, respectively, where we base ourselves on Solar System knowledge.

After the first 1~Myr the system reaches steady state. Also, in the first several Myr the HR~8799 belts develop orbital structure with gaps caused by the mean motion resonances with the planets (confirming previous work by \citet{Contro2016} and \citet{Read2018}). All four planets experience minor body impacts from the inner and outer belts. The innermost planet HR~8799~e is affected the most by the objects from the inner belt. In turn, the outermost planet HR~8799~b experiences the most impacts from the outer belt. Our numerical integrations of the dust populations in the inner and outer belts show that there is no significant contribution from these populations to the material delivery to the giant planets. The outer belt delivers an order of magnitude more of the volatile material to the planets. The refractory delivery rates from the inner and the outer belt are similar, within the error bars.

We expect the four giant planets to be born volatile-rich, so volatile delivery through impacts is probably insignificant in comparison. The enrichment in refractory material, however, may well be significant. JWST-MIRI observations targeting silicate features could be especially diagnostic and should be considered.

%__________________________________________________________________

\section*{Acknowledgements}

We thank an anonymous referee for providing us with helpful comments that allowed us to improve and clarify the manuscript. We are thankful to \emph{Hanno Rein} for valuable help with REBOUND, \emph{Jonathan Horner} for input on the inner belt, \emph{Matthew Read} and \emph{Mark Wyatt} for input on the outer belt. Simulations in this paper made use of the REBOUND code which can be downloaded freely at \url{http://github.com/hannorein/rebound}. We would like to thank the Center for Information Technology of the University of Groningen for their support and for providing access to the Peregrine high performance computing cluster. PP's work was supported by NASA's ISFM award.

\bibliographystyle{aa} % style aa.bst
\bibliography{mybibfile} % your references Yourfile.bib

\end{document}